\documentclass{SciPost}

\hypersetup{
    colorlinks,
    linkcolor={red!50!black},
    citecolor={blue!50!black},
    urlcolor={blue!80!black}
}

\usepackage[bitstream-charter]{mathdesign}
\DeclareSymbolFont{usualmathcal}{OMS}{cmsy}{m}{n}
\DeclareSymbolFontAlphabet{\mathcal}{usualmathcal}

\fancypagestyle{SPstyle}{
\fancyhf{}
\lhead{\colorbox{scipostblue}{\bf \color{white} ~SciPost Physics }}
\rhead{{\bf \color{scipostdeepblue} ~Submission }}

\fancyfoot[C]{\textbf{\thepage}}
}

\usepackage{bm}

\begin{document}

\pagestyle{SPstyle}

\begin{center}{\Large \textbf{\color{scipostdeepblue}{
Transitions to super-radiance in ensembles of incoherently pumped emitters\\
}}}\end{center}

\begin{center}\textbf{
Laszlo Rassaert \textsuperscript{1$\star$} and
Tommaso Roscilde \textsuperscript{1$\dagger$}
}\end{center}

\begin{center}
{\bf 1} Univ Lyon, Ens de Lyon, CNRS, Laboratoire de Physique, F-69342 Lyon, France
\\[\baselineskip]
$\star$ \href{mailto:email1}{\small laszlo.rassaert@ens-lyon.fr}\,,\quad
$\dagger$ \href{mailto:email2}{\small tommaso.roscilde@ens-lyon.fr}
\end{center}

\section*{\color{scipostdeepblue}{Abstract}}
\textbf{
%
Super-radiance is a striking phenomenon resulting from the collective interaction of emitters with light, and it is fundamentally related to the appearance of long-range correlations between the dipole moments of the emitters. As such, it represents a distinct phase of dissipative many-body systems compared to the case of independent emitters, similarly to how ferromagnetism stands out in magnetic materials in contrast to paramagnetism. In this work we address the conditions under which the transition to super-radiance can occur in ensembles of emitters subject to dephasing and individual decay -- a situation which is particularly relevant to the case of emitters in the solid state. The simplifying assumption of an ensemble of permutationally invariant emitters allows for the efficient solution of both the dissipative dynamics after a pulsed excitation, as well as of the steady state under incoherent pumping. This exact solution allows us to benchmark a truncated cumulant expansion approach, which can give predictions for arbitrarily big system sizes. We show that super-radiance is fundamentally robust to sizable dephasing and individual decay rates, both under a pulsed excitation, as well as under continuous pumping. This robustness is the result of the collective acceleration effect of super-radiant emission with respect to the individual coupling to a local environment.  We establish the universal critical scaling laws at the transition between normal radiance and super-radiance; and we show that, in the super-radiant phase, significant finite-size crossovers can be observed before reaching the asymptotic scaling regime. Our results pave the way for future experiments to provide a quantitative characterization of the scaling properties of super-radiance, seen as a distinct non-equilibrium many-body phase in ensembles of incoherently pumped emitters. 
}


\vspace{\baselineskip}

\vspace{10pt}
\noindent\rule{\textwidth}{1pt}
\tableofcontents
\noindent\rule{\textwidth}{1pt}
\vspace{10pt}
\newpage

\section{Introduction and main results}

The engineering of many-body phases of matter is an overarching subject of modern quantum science. A significant effort is currently invested in the quantum simulation of many-body Hamiltonians, using synthetic many-body systems made of cold atoms \cite{GrossB2017,BrowaeysL2020}, superconducting circuits \cite{Houck2012,Juanjobook} or semiconducting nanostructures \cite{Hensgens2017,Donnelly2026}. Hamiltonian quantum simulation of quantum many-body physics is aimed at investigating diverse phenomena such as the low-energy equilibrium physics of quantum spin lattices, or of strongly correlated fermions and bosons; or at realizing transient non-equilibrium quantum many-body states with special entanglement content.  

A fundamentally alternative scheme to Hamiltonian quantum simulation is offered by dissipative quantum-state engineering, \emph{i.e.}, based on the use of the coupling of a system to its environment \cite{Verstraete2009,Harrington2022}. In this respect the \emph{collective} coupling of ensembles of emitters to their environment -- leading to phenomena such as super-radiance \cite{dicke_coherence_1954,gross_superradiance_1982,benedict_super-radiance_2018} and sub-radiance \cite{dicke_coherence_1954,DeVoeetal1996,guerin_subradiance_2016,Feriolietal2021} -- offers a unique opportunity for the preparation of many-body phases of matter, with deeply complementary strengths compared to Hamiltonian engineering. Dissipative many-body systems can realize interesting many-body states as \emph{steady states}, and not as transient effects. And such states may have no analog in the thermal equilibrium physics of Hamiltonian systems. Moreover, in the specific case of ensembles of light emitters, the state of the emitter ensemble has important repercussions on the quantum state of the emitted light -- both in the stationary regime as well as in the transient one. The outgoing light offers a direct form of diagnostics of the many-body phases; and the control on the dissipative process of emission offers a tool to engineer non-classical states of light \cite{PorrasC2008}. Finally, since the coupling to the environment is an intrinsic element of the system, the many-body physics realized with it can be in principle immune to dephasing or dissipation. Yet this last point needs to be ascertained explicitly, since the collective coupling of the emitters to a common light mode can be in competition with the individual coupling of each emitter to its local environment. This is precisely the central topic of the present study. 

In this work we focus on super-radiance, i.e. the emission of light by an array of long-range correlated dipoles coupled to a common mode. Super-radiance has been investigated in a variety of different platforms, from atomic ensembles \cite{skribanowitz_observation_1973, gross_observation_1976, gibbs_single-pulse_1977, Feriolietal2021-2, Ferioli2023, ferioli_emergence_2024, bach_emergence_2024, liedl_observation_2024} to solid-state structures \cite{raino_superfluorescence_2018, findik_high-temperature_2021,biliroglu_room-temperature_2022}; and it is theoretically well understood in its idealized setting \cite{gross_superradiance_1982, benedict_super-radiance_2018}. As for any many-body phase, its strict quantitative definition must involve well-defined scaling laws: of its correlation properties in the case of steady state super-radiance; as well as of its dynamical properties in pulsed super-radiance. In particular, the scaling properties of correlations and fluctuations are essentially analogous to those of a phase of matter displaying long-range order, such as ferromagnetism in condensed matter. The observation of the fundamental scaling properties of super-radiance in experiments remains challenging, although recent efforts in this direction are rather promising \cite{raino_superfluorescence_2018,biliroglu_room-temperature_2022}. 

The robustness of super-radiance -- namely of its fundamental scaling properties -- to further environmental effects remains poorly understood. A very important effect, hindering correlations and hence super-radiance, is the coupling of the emitters to their local environments. This is very important in the solid state, in which emitters are in principle coupled to a dephasing thermal bath; and it is generally important when emission can occur not only in a common light mode, but also in other light modes or non-radiative decay channels to which the emitters can couple individually. 

In this work we offer a thorough theoretical analysis of the conditions for the onset of super-radiance in ensembles of identical emitters enjoying permutational invariance. This fundamental simplifying assumption offers the possibility to efficiently solve the Lindblad equation of the system far beyond the range of sizes accessible to exact diagonalization in the absence of permutational symmetry, i.e. up to $N\sim 10^3 - 10^4$ emitters (instead of $N\sim 10$ in generic systems). Moreover the existence of an exact solution for such large system sizes allows us to benchmark the truncated cumulant expansion (TCE) approach for this problem \cite{ritsch_benchmarking_2025}, which turns out to be generally rather accurate, when limiting one's attention to low-order correlation properties. Importantly, this latter approach allows for an extension of the theoretical predictions up to arbitrary sizes, since the size only enters as a parameter in the TCE equations. 
Equipped with these powerful tools, we can offer a systematic study of  the robustness of super-radiance to the main independent forms of coupling to a local environment for two-level systems, \emph{i.e.}, individual dephasing and individual decay.

\begin{figure}[ht!]
\begin{center}
\includegraphics[width=\textwidth]{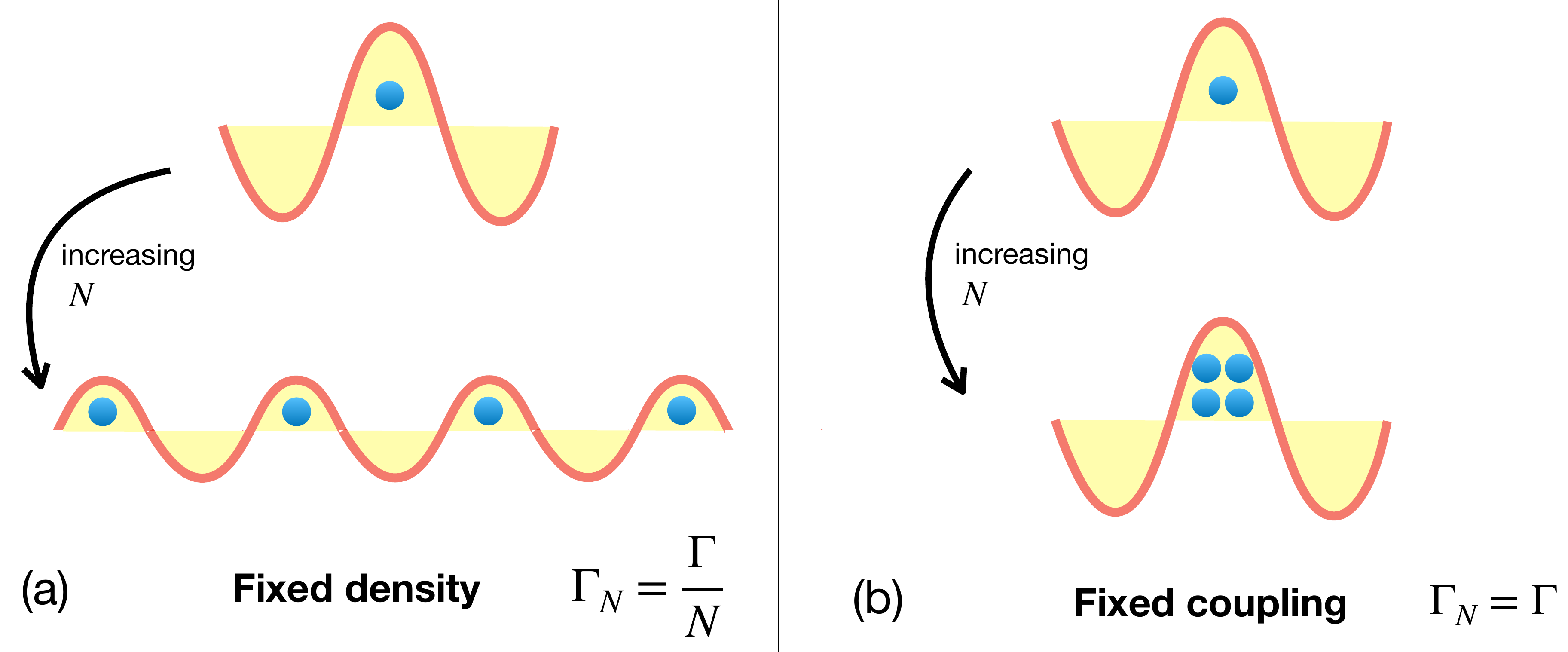}
\caption{Coupling schemes of emitters to light considered in this work: (a) \emph{fixed density}: when increasing the number of emitters, the coupling strength to the common emission mode decreases, so that the collective emission rate decreases as $\Gamma_N = \Gamma/N$; (b) \emph{fixed coupling}: when increasing the number of emitters, the coupling to the common emission mode remains the same, $\Gamma_N = \Gamma$.}
\label{f.schemes}
\end{center}
\end{figure}

Our main result is that the defining scaling properties of super-radiance can be robust to both individual dephasing and individual decay.  The precise assessment of the robustness of its scaling properties requires specifying under which conditions the number of emitters is scaled. Two situations are considered in this work -- see Fig.~\ref{f.schemes} for an illustration: 
\begin{enumerate}
\item the condition of \emph{fixed density} of emitters within the volume of the common mode. This implies that increasing the number of emitters requires increasing the volume of the mode, and hence reduce the coupling of the emitters to the mode; 
\item the condition of \emph{fixed coupling} to the common mode, independent of the number of emitters; this implies that an increasing number of emitters can be coupled to the same common mode in a symmetric fashion, without having to alter the mode geometry and hence the couplings of the emitters to it.  
\end{enumerate}

 In the case of fixed density, steady state super-radiance under incoherent pumping is found to be robust to both dephasing and individual decay, namely it takes a critical dephasing rate and a critical individual decay rate to observe a dissipative phase transition from super-radiance to collective radiance with only effective short-range correlations. This observation has also significant implications for the case of fixed coupling: regardless of their strength, dephasing and individual decay will always be overcome by collective decay upon increasing the system size.

 Regarding the case of pulsed super-radiant emission, and for a fixed density of emitters, super-radiant decay is found to be robust to dephasing up to a critical rate. On the other hand, individual decay introduces a critical size $N_c$ beyond which the super-radiant scaling of the emission intensity is lost. Interestingly, though, the critical size diverges exponentially when the individual decay rate goes to zero. When considering instead a fixed coupling of the emitters to the common mode, dephasing and individual decay are systematically overcome by collective emission upon increasing the system size, and super-radiant scaling is robust for systems exceeding a certain threshold in size. 
 
 Moreover we establish universal features of the transition from super-radiant to normal scaling of correlations, both in the case of super-radiant bursts as well as in the case of steady state super-radiance. In the steady state case, the super-radiant transition is accompanied by universal critical properties in the photon statistics of the emitted light. On the other hand, the onset of a super-radiant burst in the presence of dephasing is accompanied by critical slowing-down of the emission, and by unconventional scaling of the peak intensity.   
 
 The main results of our work on the stability of super-radiance are summarized in the two tables, \ref{tab:fixed_density} and \ref{tab:fixed_coupling}. 
 They establish that super-radiance and its fundamental scaling properties can be observed under very general conditions in systems which can feature high densities of emitters, such as in solid-state nano-structures (quantum dots, quantum wells) \cite{raino_superfluorescence_2018, findik_high-temperature_2021,biliroglu_room-temperature_2022}. The super-radiant emission in these systems can be robust to a thermal environment up to potentially high temperatures (the higher the larger the number $N$ of emitters which can be coupled symmetrically to the common mode); and it is robust as well to the presence of individual (such as non-radiative) decay processes.

   \begin{table*}[ht!]
    \caption{Robustness of super-radiance phenomena to the presence of a local environment in the case of \emph{fixed density} of the emitters inside the volume of the common emission mode.}
    \label{tab:fixed_density}
  \begin{center}
   \begin{tabular}{l||l|l|} 
      {\bf Fixed density} & individual dephasing ($\gamma_\phi$) & individual decay  ($\gamma_d$)\\
    \hline
    \hline
      steady state super-radiance &  survives up to a critical  $\gamma_\phi$ &  survives up to a critical $\gamma_d$ \\
      \hline
      pulsed super-radiance & survives up to a critical  $\gamma_\phi$  & \begin{tabular}{llr} survives up to a critical \\ number of emitters ~~~~~~\\ $N_c \sim O[\exp(1/\gamma_d)]$ ~~~~~\end{tabular} \\
       \hline
        \end{tabular}
  \end{center}
\end{table*}

 \begin{table*}[ht!]
  \caption{Robustness of super-radiance phenomena to the presence of a local environment in the case of a \emph{fixed coupling} of the emitters to the common emission mode.}
       \label{tab:fixed_coupling}
  \begin{center}
    \begin{tabular}{l||l|l|} 
      {\bf Fixed coupling} & individual dephasing ($\gamma_\phi$) & individual decay  ($\gamma_d$)\\
         \hline
         \hline
      steady state super-radiance &  \begin{tabular}{llr} survives for any  $\gamma_\phi$ \\ (with $\gamma_p = O(N)$)
      \end{tabular} &  \begin{tabular}{llr} survives for any $\gamma_d$ \\ (with $\gamma_p = O(N)$) \end{tabular} \\
      \hline
      pulsed super-radiance & \begin{tabular}{llr} survives for any $\gamma_\phi$ \\ (threshold in $N$) \end{tabular} &  \begin{tabular}{llr} survives for any $\gamma_d$ \\  (threshold in $N$) \end{tabular} \\
       \hline
    \end{tabular}
      \end{center}
\end{table*}

The structure of the paper is as follows. Sec.~\ref{s.model} presents the model and Sec.~\ref{s.method} the method and approximation schemes we used. Sec.~\ref{s.steady_state_results} describes the super-radiant phase transition under incoherent pumping and the analytical results obtained with our approximations, while Sec.\ref{s.numerical_results} presents exact results obtained with numerical methods. Finally, Sec.\ref{s.pulsed_super-radiance} presents numerical results for the pulsed super-radiance in the presence of local environment.

\section{Permutationally invariant emitters under incoherent pumping}
\label{s.model}
\subsection{Model}

In this work we consider ensembles of incoherently pumped emitters subject to collective emission (CE) into a common mode; as well as to individual decay (ID) into different modes and to dephasing. The dynamics is assumed to be purely driven by the coupling to a Markovian environment, and it is therefore described by the Lindblad equation: 

\begin{equation}
   \dfrac{d \rho}{d t}  = \mathcal{L}_p\left[ \rho \right] + \mathcal{L}_{\rm CE} \left[ \rho \right] + \mathcal{L}_{\rm ID} \left[ \rho \right] +  \mathcal{L}_\phi \left[ \rho \right] 
   \label{e.lindblad}
\end{equation}

where

\begin{eqnarray}
    \mathcal{L}_p\left[ \rho \right] & = &  \gamma_p \sum_{i = 1}^{N} \left( S_i^+ \rho S_i^- - \frac{1}{2} \left\{ S_i^- S_i^+, \rho \right\} \right) \nonumber \\
    \mathcal{L}_{\rm CE} \left[ \rho \right] & = &  \Gamma_N \sum_{i, j = 1}^{N}  \left( S_i^- \rho S_j^+ - \frac{1}{2} \left\{ S_j^+ S_i^-, \rho \right\} \right) \nonumber \\
    \mathcal{L}_{\rm ID}  \left[ \rho \right] & = & \gamma_{d} \sum_{i = 1}^{N}  \left( S_i^- \rho S_i^+ - \frac{1}{2} \left\{ S_i^+ S_i^-, \rho \right\} \right)  \nonumber \\
    \mathcal{L}_\phi \left[ \rho \right] & = & \gamma_\phi \sum_{i = 1}^{N}  \left( S_i^z \rho S_i^z - \frac{1}{4} \rho  \right)~.
    \label{e.lindbladians}
\end{eqnarray}
Here $S_i^\mu$ $(\mu = +, -, z)$ are spin-1/2 operators associated with the $i$-th emitter. 

The incoherent pumping term ${\cal L}_p$ describes a process of incoherent excitation of the emitters with a rate $\gamma_p$. In the context of \emph{e.g.}, solid-state nanostructures, such as quantum dots \cite{biliroglu_room-temperature_2022, findik_high-temperature_2021, raino_superfluorescence_2018}, the emitters are typically driven by an applied field which is blue-shifted with respect to their emission frequency; hence, before decaying, the emitter must lose its extra energy by non-radiative transitions involving e.g. phonons. Hence, even if the driving field is a coherent laser, its phase coherence is lost  in the intermediate transitions. 

The collective emission term ${\cal L}_{\rm CE}$ describes the spontaneous decay at a rate $\Gamma_N$ into a common mode for which the emitters are indistinguishable -- either because 1) they are located at equivalent positions with respect to the mode profile, as sketched in Fig.~\ref{f.schemes}(a); or because 2) they are all contained within a $\lambda^3$ volume ($\lambda$ being the mode wavelength), as sketched in Fig.~\ref{f.schemes}(b). The collective emission rate $\Gamma_N$ is size-dependent in the first case (fixed density), and it takes the form $\Gamma_N = \Gamma/N$, since increasing the size of the system implies increasing the volume of the mode and hence decreasing the coupling to each emitter. On the other hand, the collective emission rate becomes size-independent in the second case (fixed coupling), $\Gamma_N = \Gamma$. 

The individual emission term ${\cal L}_{\rm ID}$ describes the decay of the emitters at a rate $\gamma_d$ due to processes other than the collective emission one -- \emph{e.g.}, non-radiative decay processes for solid-state emitters \cite{Saleh_Teich}, or emission into other electromagnetic modes. 

Finally the dephasing term ${\cal L}_{\phi}$ describes the fluctuations of the interactions of the emitters with their local environment, altering their emission frequency and causing therefore dephasing at a rate $\gamma_\phi$. This effect can come from non-resonant coupling of the emitters to \emph{e.g.}, thermally excited phonons. 

 More generally, the terms in Eq.~\eqref{e.lindbladians} comprise nearly all possible mechanisms of interaction of an ensemble of $S=1/2$ spins with an environment, which exhibit \emph{permutational invariance} \cite{shammah_open_2018, xu_simulating_2013}. In fact, we excluded two further possible processes, namely collective pumping and collective dephasing, which are in general not realistic for the systems of our interest. 
 
 The permutational invariance of the system of our interest leads to the possibility of achieving an exact solution to the Lindblad equation for much larger systems compared to the case of generic, \emph{i.e.}, non-symmetric systems \cite{xu_simulating_2013, shammah_open_2018}. In this work we solve exactly systems with up to $N=5\times 10^3$ emitters. We describe briefly in the following Sec.~\ref{s.PI} how a solution can be obtained efficiently under permutational invariance. On the other hand, we observe that in some regimes the scaling of correlations has not reached its asymptotic behavior even for the largest sizes accessible to the exact solution. Hence, for those cases we supplement the exact solution with a truncated cumulant expansion up to second order \cite{Colussietal2018, Sanchez-Barquilla2020, ritsch_benchmarking_2025}, described in Sec.~\ref{s.TCE2}. This approach turns out to be accurate in the description of two-point correlations for the sizes covered by the exact solution. Therefore it can extend our predictions to arbitrary system sizes, revealing the asymptotic scaling behavior. 
 
\subsection{Exact solution exploiting permutational invariance}
\label{s.PI}

In both configurations considered in Fig.~\ref{f.schemes}, the quantum emitters are indistinguishable for the light mode. This symmetry - invariance under permutation -  makes it possible to re-write the problem in term of collective spin variables only, ${\bm J} = \sum_{i=1}^N {\bm S}_i$, and their eigenstates \cite{xu_simulating_2013, shammah_open_2018}. We focus in particular on the joint eigenstates of the collective-spin operators ${\bm J}^2$ and $J^z$, $ | J, M, \lambda \rangle$ where ${\bm J}^2 = J(J+1)$ and $J = \{ N/2, N/2-1, ..., 0 \}$, while $J^z = M$ and $M = \{ -J, ..., J \}$. $\lambda$ is a set of additional quantum numbers necessary to uniquely label the states for $J< N/2$. The $2^N$-dimensional Hilbert space is then broken up into $O(N^2)$ sectors labeled by the quantum numbers $(J,M)$, each sector containing 
\begin{equation}
    \label{e.Djm}
    D(J,M) = (2J +1) \frac{N!}{(\frac{N}{2} + J + 1)! (\frac{N}{2} - J)!}
\end{equation}
states \cite{arecchi_atomic_1972}. 
In the absence of any Hamiltonian term, the evolution induced by the dissipators in Eq.~\eqref{e.lindbladians} cannot induce any coherence between Dicke states with different $J,M$ indices in the density matrix. Hence, unless coherence is present in the initial state of the evolution, the density matrix is block diagonal with blocks labeled by $J, M$; and this condition will always be met by the steady state of the system. We shall make the assumption of a block-diagonal structure in the rest of this work.  
 Moreover all matrix elements between states in the same sector are equivalent because of permutational invariance, \emph{i.e.}, $\langle JM\lambda | \rho | JM\lambda' \rangle = \delta_{\lambda, \lambda'} \rho_{JM} /  D(J, M)$. This implies that, for all the situations of interest to this work, the $|JM\lambda\rangle$ basis is the density-matrix eigenbasis, and each $(J,M)$ block is proportional to the identity, $\mathbb{1}_{D(J,M)\times D(J,M)}$. 
 
 
 Therefore the density matrix is known by just knowing $\sum_{J=0}^{N/2} (2J+1) = (N/2+1)^2$ matrix elements $\rho_{JM}$. Hence it becomes numerically accessible to solve exactly the Lindblad equation for rather large systems. In Sec.~\ref{s.numerical_results}, we show numerical results for system size up to $N = 5 \times 10^3$ emitters for the steady state solution, and $N = 10^3$ for the dynamics of the system in Sec.~\ref{s.pulsed_super-radiance}. 

\section{Equations of motion and approximate treatments}
\label{s.method}

In permutationally invariant ensembles of emitters, the only meaningful observables of interest are associated with the collective spin ${\bm J} = \sum_{i=1}^N {\bm S}_i$. We shall characterize the various regimes of open-system dynamics by making use of first- and second-order moments of the fluctuations of the collective spin. Be $N_e$ the number of excited emitters, given by $N_e = J^z +N/2$.

The exact steady state of the Lindblad equation Eq.~\eqref{e.lindblad} does not admit any coherence between states with different numbers of excited emitters (or different $J^z$ magnetization), namely the average value of any operator not conserving $J^z$ is zero, \emph{e.g.}, $\langle J^+ \rangle = \langle J^- \rangle = 0$, $\langle J^+ J^+ \rangle = \langle J^- J^-\rangle = 0$, etc. This is due to the fact that the emitter ensemble is incoherently pumped: any coherence induced in the initial state will decay under the effect of the coupling to the environment.
Moreover we shall consider transient evolutions starting from initial states in which no coherence is imprinted. As a consequence, no coherence will develop either during the evolution. The vanishing average of  all operators not conserving $J^z$ will therefore extend to all of our calculations (except for an approximation scheme neglecting second-order cumulants, see below). 
In the following we restrict our focus to two $J^z$-conserving quantities built from collective-spin operators: $J^z$ itself, and the integrated dipole-dipole correlations $J^+ J^- = \sum_{ij} S_i^+ S_j^-$, which, as we will see further, are related to the intensity of the emitted light.

The equations of motion for the average values of these two quantities under the Lindblad evolution, namely $d\langle A \rangle/dt  = {\rm Tr}(A ~d\rho/dt)$, read  

\begin{eqnarray}
    \frac{d \langle J^z \rangle}{dt} &=&  \gamma_{p} \left( \frac{N}{2} -  \langle J^z \rangle \right) 
    - \gamma_d \left( \frac{N}{2}+ \langle J^z \rangle \right) - \Gamma_N \langle J^+ J^- \rangle 
    \label{e.dJzdt}
    \\
    \frac{d \langle J^+ J^- \rangle}{dt} &=& 
    \gamma_p \left( N - \langle J^+ J^- \rangle \right)   
    - \gamma_d \langle J^+ J^- \rangle
    +  2 \Gamma_N \langle J^+ J^z J^- \rangle   
    + \gamma_\phi \left(  \frac{N}{2} - \langle J^+ J^- \rangle + \langle J^z \rangle \right) ~~~
    \label{e.dJpJmdt}
\end{eqnarray}

As one can observe, $\Gamma_N \langle J^+ J^- \rangle$ provides the speed of decay of the excited-state population $d\langle N_e \rangle/dt = d \langle J^z \rangle /dt$ due to collective emission, and thus it is equivalent to the intensity of the emitted light, which is the key element for super-radiance. The equation for the evolution of the dipole-dipole correlations  $\langle J^+ J^- \rangle$ contains the term $\langle J^+ J^z J^- \rangle$, whose equation of motion should also be established, and so on -- generating in this way a hierarchy of $O(N^2)$ coupled equations. 

A well-established approximation schemes to treat these equations consists in truncating the hierarchy, so as to obtain a straightforward (yet approximate) solution.
The idea of the truncation scheme is to neglect cumulants of the spin fluctuations starting from a given order \cite{Colussietal2018, Sanchez-Barquilla2020, ritsch_benchmarking_2025}. Cumulants of the fluctuations associated with the density matrix are defined recursively as 
\begin{itemize}
\item first order: $\langle A \rangle = \langle A \rangle_c$
\item second order:  $\langle A B\rangle = \langle A B \rangle_c + \langle A \rangle_c \langle B \rangle_c$
\item third order: $\langle A B C \rangle = \langle A B C \rangle_c + \langle A  B \rangle_c \langle C \rangle_c
+ \langle A  C \rangle_c \langle B \rangle_c   \nonumber + \langle B C \rangle_c \langle A \rangle_c
+ \langle A \rangle_c  \langle B \rangle_c \langle C \rangle_c  $
\item ...
\end{itemize}

\subsection{Truncated cumulant expansion (TCE) to first order} 
\label{s.TCE1}
The truncation of the cumulant hierarchy to lowest order assumes that all cumulants vanish starting from second order - we will call this approximation TCE1 in the following. It then implies $\langle J^+ J^- \rangle \approx \langle J^+ \rangle \langle J^- \rangle$, so that one is obliged to assume a finite coherence between different $J^z$ sectors, \emph{i.e.}, $\langle J^- \rangle \neq 0$, for collective emission to be relevant, even though the exact results show no such coherence. The equation of motion for the first-order coherence term is: 
\begin{equation}
\frac{d \langle J^+ \rangle}{dt}  =  \Gamma_N \langle J^+ J^z \rangle  - \frac{1}{2} \left ( \gamma_p + \gamma_d + \gamma_\phi \right) \langle J^+ \rangle 
\approx  \left [  \Gamma_N  \langle  J^z \rangle  - \frac{1}{2} \left ( \gamma_p + \gamma_d + \gamma_\phi \right) \right ]  \langle J^+ \rangle ~.
\label{e.dJpdt}
\end{equation}

The approximation neglecting second-order cumulants is akin to a mean-field one for the density matrix, although strictly speaking this is only true when $N \to \infty$. Indeed a mean-field approximation, i.e. a factorized Ansatz for the density matrix in the form $\rho =: \otimes_{i=1}^N \rho_i$, implies the factorization of the correlation function between different spins, namely  $\langle S_i^\mu S_j^\nu \rangle =:  \langle S_i^\mu \rangle \langle S_j^\nu \rangle$, which, under permutation invariance, entails $\langle J^\mu J^\nu \rangle =: \langle J^\mu \rangle \langle J^\nu \rangle (1- 1/N)+ \frac{i}{2} \sum_\gamma \epsilon_{\mu\nu\gamma} \langle J^\gamma \rangle$. The second term in the previous expression, being $O(N)$, is a correction to the dominant term  $\langle J^\mu \rangle \langle J^\nu \rangle \sim O(N^2)$ (when this term is finite). This refined approximation is detailed in Appendix \ref{a.TCEb}.

The physics of all-to-all interacting systems is generally thought to be well captured by a mean-field approximation. Although the latter correctly predicts certain aspects of the phase diagram, we will see that moving beyond the TCE1 scheme allows one to describe important aspects which are completely missed at the mean-field level, notably the critical scaling at the phase transition between the super-radiant and the normal phase.

\subsection{Truncated cumulant expansion to second order}
 \label{s.TCE2} 
 
As we shall see in the following, a much more effective approximation consists in retaining second-order cumulants, while postulating the vanishing of third- and higher-order cumulants, namely 
$\langle A B C \rangle_c = 0$. We shall call this approximation TCE2 in the following. While the truncation of cumulant expansion to first order (Sec.~\ref{s.TCE1}) corresponds to a mean-field Ansatz on the state of the system,  the assumption of vanishing cumulants beyond second order is not verified by any explicit quantum spin state. Yet it leads to a very practical closure of the system of Eqs.~\eqref{e.dJzdt} and \eqref{e.dJpJmdt}. Indeed in its simplest formulation the TCE2 approach posits that 
\begin{equation}
\langle J^+ J^z J^- \rangle =: \langle J^+ J^- \rangle \langle J^z \rangle \label{e.TCE2}
\end{equation} 
when retaining only the magnetization-conserving terms. A more refined version of this approximation consists in applying it uniquely to three-point correlators between different spins, namely writing $\langle S_i^+ S_j^z S_l^- \rangle =: \langle S_i^+ S_l^- \rangle \langle S_j^z \rangle$ for $i \neq j \neq l$. This refined version is detailed in Appendix \ref{a.TCEb}, but is not used to obtain the results of this paper, as it only adds subleading terms in the equations.

\section{Steady state under continuous incoherent pumping} 
\label{s.steady_state_results}

We first focus on the steady state solution of the Lindblad dynamics dictated by Eq.~\eqref{e.lindblad}, and in particular on its correlation properties marking the different phases of the stationary regime. 

\subsection{Definition of super-radiant, normal and sub-radiant phases} 

Here and in the rest of this work, a \emph{super-radiant} (SR) phase will be strictly characterized by long-range correlations between the dipoles of the emitters, namely by the scaling property
\begin{equation}
{\rm SR:} ~~ \langle J^+ J^ - \rangle  = \sum_{ij} \langle S_i^+ S_j^- \rangle \sim O(N^2)~. 
\end{equation}
Given that in our permutationally invariant systems $ \langle J^+ J^ - \rangle = N(N-1) C^{+-} + \langle J^z \rangle + \frac{N}{2}$, with $C^{+-} = \langle S_i^+ S_j^- \rangle_{i \neq j}$ independent of the choice of sites $i,j$, the condition for having super-radiance is that $C^{+-}\sim O(1)$, which amounts to finite dipole-dipole correlations for all pairs of dipoles.  In the following, $\langle J^+ J^- \rangle / N^2$ will play the role of the square of the order parameter for the super-radiant phase transition.

The opposite limit is that of vanishing  correlations $C^{+-} = 0$, where $ \langle J^+ J^ - \rangle = \langle J^z \rangle + \frac{N}{2} = \langle N_e \rangle$, as in a system of completely independent emitters. Any intermediate case between  $ \langle J^+ J^ - \rangle  = \langle N_e \rangle$ and  $ \langle J^+ J^ - \rangle \sim O(N^2)$ is a manifestation of correlated emission, i.e. 
$C^{+-} > 0$. Yet one could have $C^{+-} \to 0$ when $N\to \infty$, so that finite correlations do not survive for large systems. We will generally qualify all regimes in which $ \langle J^+ J^ - \rangle  \sim O(N)$ as \emph{normal}, even though they may manifest correlated emission properties. Critical regimes separating super-radiant from normal ones can instead be characterized by an anomalous scaling, \emph{i.e.}, $ \langle J^+ J^- \rangle \sim O(N^{1+\epsilon})$ with $0 < \epsilon < 1$.    

A further possibility is that emitters anti-correlate, i.e. $C^{+-} <0$, so that $ \langle J^+ J^ - \rangle$ takes a smaller value than in the case of uncorrelated emitters, namely $ \langle J^+ J^ - \rangle  <\langle N_e \rangle$. Such a behavior marks a \emph{sub-radiant} phase.

In the absence of coherence between different magnetization sectors, the quantity 
\begin{equation}
\langle J^+ J^- \rangle = \langle (J^x)^2  + (J^y)^2 + J^z \rangle = {\rm Var}(J^x) + {\rm Var}(J^y) + \langle J^z \rangle
\end{equation} 
(since $\langle J^x\rangle = \langle J^y \rangle = 0$) captures the fluctuations of the $x$ and $y$ components of the collective spins. In the presence of long-range correlations for the $x$ and $y$ spin components, $\langle J^+ J^- \rangle$ is dominated by the  ${\rm Var}(J^{x(y)})$ contributions scaling as  $\sim O(N^2)$, namely
\begin{equation}
\langle J^+ J^- \rangle =  [{\rm Var}(J^x) + {\rm Var}(J^y)] (1+ O(1/N))~. 
\end{equation}
In magnetic systems, the collective-spin variances ${\rm Var}(J^{x(y)})/N$ go under the name of (static) structure factors, growing as $N$ in the presence of long-range order. In an ensemble of emitters, the quantity $\langle J^+ J^- \rangle$ acquires the further significance of collective emission rate (see Eq.~\eqref{e.dJzdt}), given by $\Gamma_N \langle J^+ J^- \rangle$; but also that of intensity of the emitted electromagnetic field, when using the correspondence $E^{(\pm)} \sim J^{\mp}$ between the spin operators and the electric field operators.

\begin{figure}[ht!]
\begin{center}
\includegraphics[width=\textwidth]{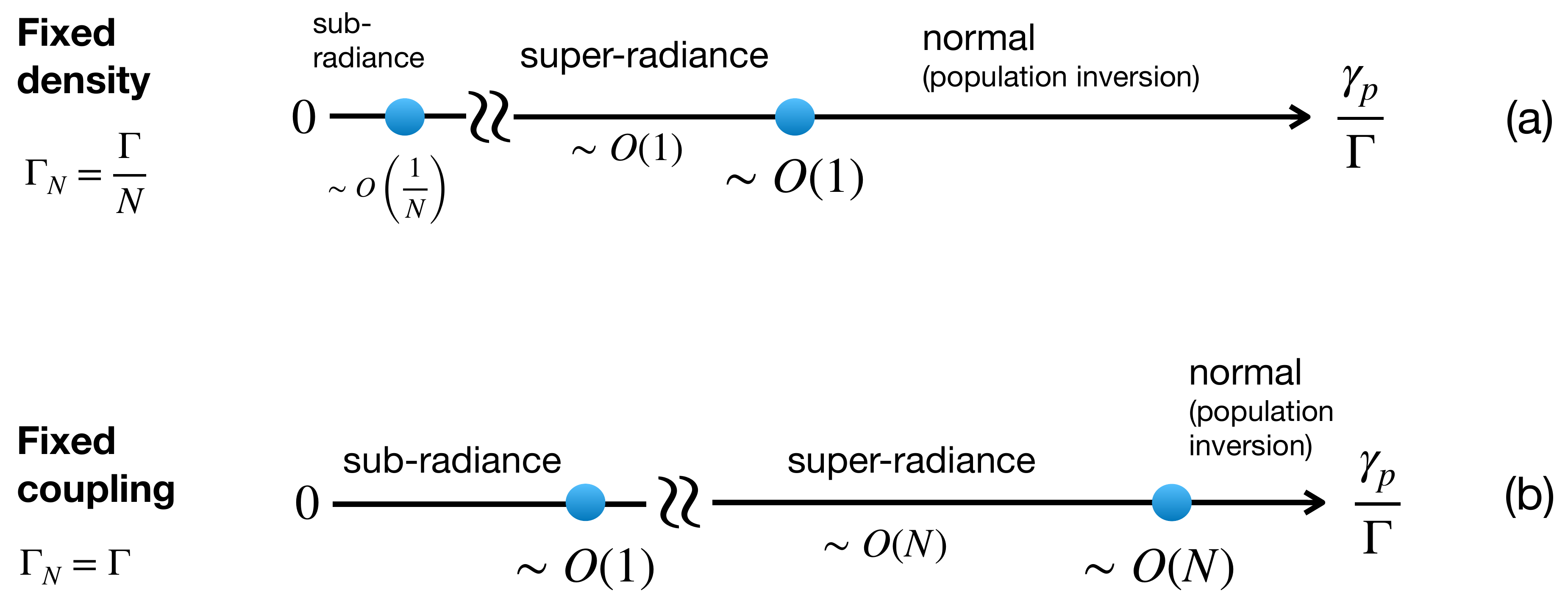}
\caption{Steady-state phase diagrams within the two coupling schemes of the emitters to the common emission mode: (a) fixed-density scheme; (b) fixed-coupling scheme.}
\label{f.steady_PhD}
\end{center}
\end{figure}

\subsection{Super-radiant-to-normal transition, and sub-radiant-to-super-radiant transition}
\label{s.SR_TCE1}

Our goal is to investigate the competition between normal and super-radiant phases in the steady state solution of the Lindblad equation Eq.~\eqref{e.lindblad}, due to the competition between the local dissipation/dephasing/pumping terms on the one side, and the collective emission term on the other.  In the absence of dephasing and individual decay, a super-radiant-to-normal phase transition has already been observed in Refs.~\cite{meiser_prospects_2009,meiser_intensity_2010, meiser_steady-state_2010, shammah_superradiance_2017, shankar_subradiant--subradiant_2021}. Its existence can be grasped at the level of the truncation of the cumulant hierarchy to first-order (TCE1, Sec.~\ref{s.TCE1}). Within that scheme Eqs.~\eqref{e.dJpdt} give two steady state solutions, $\langle J^+ \rangle = 0$ (normal phase) and $\langle J^+ \rangle \neq 0$ (super-radiant phase). If $\langle J^+ \rangle \neq 0$, one obtains

\begin{equation}
\langle J^z \rangle \approx \frac{N}{2}  \left ( r_p + r_d + r_\phi\right )    \\ 
\label{e.JzTCE1}
\end{equation}
where we have introduced the dimensionless rates 
\begin{equation}
r_p = \frac{\gamma_p}{\Gamma_N N}~~~~~~
r_d = \frac{\gamma_d}{\Gamma_N N} ~~~~~~
r_\phi = \frac{\gamma_\phi}{\Gamma_N N}~.
\end{equation}
When injecting Eq.~\eqref{e.JzTCE1} into Eq.~\eqref{e.dJzdt} (without the need for cumulant truncation) one obtains  
\begin{equation}
 \langle J^+ J^- \rangle  \approx   \frac{N^2}{2}  \left [ r_p - r_d  - (r_d+r_p)(r_p+r_d+r_\phi) \right] 
 \label{e.JplusJmoins_TCE1}
\end{equation}
which can serve as a basis to establish the presence of super-radiance in the steady state.

The transition from super-radiant to normal behavior occurs when the order parameter $\langle J^+ J^- \rangle/N^2$ drops to zero, \emph{i.e.}, at the critical pumping rate
\begin{eqnarray}
  r_{p,c}^{\pm} &=& \frac{1}{2}  \Big( 1 - 2 r_d - r_\phi \pm \sqrt{\left(1 - r_\phi \right)^2 - 8 r_d } \Big)  
\label{e.critical_TCE1}
\end{eqnarray}

In particular, introducing the threshold value for the individual decay rate, $r_{d,\rm th} = \frac{1}{8} (1 - r_\phi)^2 $, we find that the critical point of the super-radiant-to-normal transition admits two solutions for $r_d < r_{d,\rm th}$; one solution for $r_d = r_{d,\rm th}$; and no solution for 
$r_d > r_{d,\rm th}$.

A particularly readable case is the one in which $r_d = 0$ (no individual decay). 
In that case 
\begin{equation}
\langle J^+ J^- \rangle  \approx \frac{N^2}{2}~r_p~(1-r_p - r_\phi)~. \label{e.J+J-TCE1}
\end{equation}
This expression contains the essence of steady state super-radiance. One sees that, within the \emph{fixed density} scheme, \emph{i.e.}, $\Gamma_N N = \Gamma$ independent of $N$, the dimensionless rates are size independent,  $r_p = \gamma_p/\Gamma$ and $r_\phi = \gamma_\phi/\Gamma$. If $r_\phi < 1$, a finite pumping rate  $0< r_p < 1- r_\phi$, counterbalancing the collective emission, stabilizes super-radiance, i.e. $\langle J^+ J^- \rangle \sim O(N^2)$. On the other hand, a too strong pump $r_p \geq 1- r_\phi$ brings the system to a normal regime, associated (within the TCE1 approximation) with perfect population inversion $\langle J^z \rangle = \frac{N}{2}$. 

\begin{figure}[ht!]
\begin{center}
\includegraphics[width=\textwidth]{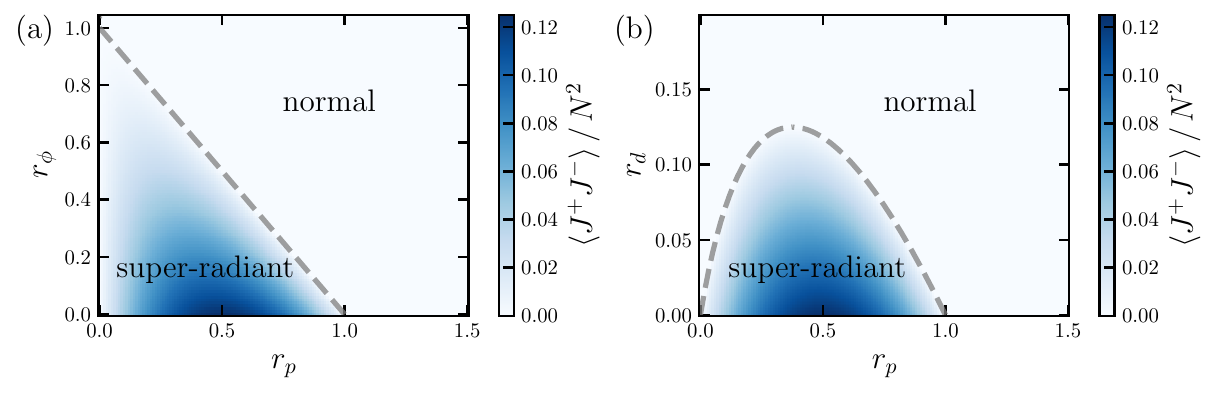}
\caption{Steady-state phase diagram in the  pumping vs. dephasing plane (a), and in the pumping vs. individual decay plane (b). The phases are characterized by the order parameter $\langle J^+ J^- \rangle / N^2$ obtained with the TCE1 approximation. The grey dashed line is the critical line given in Eq.~\eqref{e.critical_TCE1}). }
\label{f.TCE1_phase_diagram}
\end{center}
\end{figure}

We draw the phase diagram obtained within this approximation in Fig.~\ref{f.TCE1_phase_diagram}. One sees that the super-radiant phase survives both the presence of finite dephasing and of individual decay rate. These local effects simply shrink the interval of $r_p$ values over which super-radiance is stabilized in the steady state. 

Within the \emph{fixed coupling} scheme, on the other hand, the dimensionless rates are $r_p = \gamma_p/(\Gamma N)$ and $r_\phi = \gamma_\phi/(\Gamma N)$. Therefore one needs a 
pumping rate $\gamma_p$ scaling linearly with $N$ in order to stabilize super-radiance. A small, fixed pumping rate leads instead to a different regime, in which  $\langle J^+ J^- \rangle$  can be arbitrarily small, going even below the limit of uncorrelated emitters, $\langle J^+ J^- \rangle_{\rm uncorr} = \langle N_e \rangle$, which, for $N \gg 1$, is obtained when $\gamma_p/\Gamma < 1$. Under this condition, the emitters show anticorrelations and a lower intensity of emitted radiation than in the case of uncorrelated emitters, namely sub-radiance. The occurrence of sub-radiance in this system was already discussed in Ref.~\cite{meiser_steady-state_2010}.

Fig.~\ref{f.steady_PhD} summarizes the above considerations. The three steady state phases (normal, super-radiant and sub-radiant) can be observed within both coupling schemes (fixed density or fixed coupling). Yet each transition between the various phases (sub-radiant to super-radiant, super-radiant to normal) is well defined -- i.e. the critical point is size independent -- only within one of the two coupling schemes. The super-radiant-to-normal transition is well defined at a critical ratio $\gamma_p/\Gamma \sim O(1)$ only within the fixed-density scheme; while the sub-radiant-to-super-radiant transition is only defined at a critical ratio $\gamma_p/\Gamma \sim O(1)$ within the fixed-coupling scheme. In the latter scheme, a super-radiant scaling of $\langle J^+ J^- \rangle$ is only properly defined if the pumping rate $\gamma_p$ scales linearly with $N$, so that the whole ensemble can exhibit a stationary intensity of emission scaling as $O(N^2)$.

\subsection{Steady-state solution within second-order cumulant truncation}
\label{s.SR_TCE2}

In the previous section we have seen that the cumulant truncation scheme at the first-order level can capture the essential features of the steady-state phase diagram. Yet the phase transitions within that approximation scheme are completely trivial. In particular, the super-radiant-to-normal transition occurs only when the population is fully inverted, i.e. $\langle J^z \rangle = N/2$, namely the critical point is the pure state $\otimes_i |\uparrow\rangle_i$. As we will see later in the exact solution, this picture is not at all accurate, and in fact the transition point is a highly nontrivial state, with anomalous critical scaling of correlations. Interestingly, a quantitative description of the critical point can be obtained by taking  correlations between emitters into account within the cumulant truncation scheme to \emph{second} order.

Within the TCE2 scheme described in Sec.~\ref{s.TCE2}, the equations defining the steady state value for $\langle J^z \rangle$ and $\langle J^+ J^- \rangle$ are 


\begin{align}
&     r_{p} \left( \frac{N}{2} -  \langle J^z \rangle \right) 
    - r_d \left( \frac{N}{2}+ \langle J^z \rangle \right) - \frac{\langle J^+ J^- \rangle }{N} = 0 \nonumber \\
& r_{p} \left( N - \langle J^+ J^- \rangle \right)  
    - r_d \langle J^+ J^- \rangle
    +  \frac{2}{N} \langle J^+  J^- \rangle \langle J^z \rangle  
    + r_\phi \left(  \frac{N}{2} - \langle J^+ J^- \rangle + \langle J^z \rangle \right) = 0 ~.
    \label{e.dJzss}
\end{align}

Hence $\langle J^z \rangle = \frac{r_p - r_d}{r_p + r_d} \frac{N}{2} - \frac{1}{r_p + r_d} \frac{\langle J^+ J^- \rangle}{N}$ and   $X = \langle J^+ J^- \rangle$ is solution of the equation  $a X^2 - bX - c = 0$ with


\begin{align}
&    a = \frac{2}{N^2} \frac{1}{r_p + r_d} \nonumber \\
&    b =  \frac{r_p - r_d}{r_p + r_d} - r_p - r_\phi - r_d - \frac{1}{N} \frac{r_\phi}{r_p + r_d}
\nonumber \\
&    c = N \left( r_p + \frac{r_\phi}{2} \left( 1 + \frac{r_p - r_d}{r_p + r_d} \right) \right)
\end{align}

As seen in Sec.~\ref{s.SR_TCE1}, steady state super-radiance can be achieved in the fixed-density scheme ($\Gamma_N = \Gamma/N$) via a size-independent pumping $\gamma_p \sim O(1)$; or within the fixed-coupling scheme ($\Gamma_N = \Gamma$) via a linearly growing pumping rate, $\gamma_p \sim O(N)$. Within either of these schemes,    three scaling regimes for the solution can be identified upon varying the value of the $b$ coefficient: 
\begin{itemize}
    \item if $b>0$, $\langle J^+J^-\rangle \approx \frac{b}{a} \propto N^2$~ (super-radiance) 
    \item if $b=0$, $\langle J^+J^-\rangle \approx \sqrt{\frac{c}{a}} \propto N^{3/2}$ ~ (critical)
    \item if $b<0$, $\langle J^+J^-\rangle \approx \frac{-c}{b} \propto N$ ~ (normal)
\end{itemize}

In particular, the TCE2 scheme identifies a specific scaling regime corresponding to the critical point $r_{p,c}$ between the super-radiant and the normal regime, given by Eq.~\eqref{e.critical_TCE1}. Yet within the TCE2 scheme the critical point is not a trivial state with full population inversion, but a special state with anomalous critical scaling of correlations.

\subsection{Comparison with thermal transitions}


The specific scaling law emerging at the super-radiant transition point is reminiscent of thermal phase transitions for systems at equilibrium. Their critical scaling is dictated by critical exponents of the transition.
The collective nature of dissipation in our system suggests that we should compare the critical scaling with that exhibited by a system with collective U(1)-symmetric spin-spin interactions. 
The latter system sits above the upper critical dimension, and therefore its transition belongs to the mean-field universality class. 

For a mean-field phase transition occurring in a system above the upper critical dimension $d_c = 4$,  the order parameter at criticality behaves as $ \langle |m|^2 \rangle \propto N^{-2\beta / \nu d_c}$, with $\beta = 1/2$, $\nu = 1/2$, $d_c = 4$, such that $\langle |m|^2 \rangle \propto N^{- 1/2}$ \cite{Botetetal1982,BotetJ1983,Defenuetal2023}. As a consequence, one obtains 
\begin{equation}
\langle J^+ J^- \rangle = \sum_{ij} \langle S_i^+ S_j^- \rangle \sim  \langle |m|^2 \rangle N^2 \sim  N^{2-2\beta / \nu d_c} = N^{3/2}~.
\end{equation}
This coincides with the critical scaling predicted by the TCE2 approximation. 

The apparent correspondence between equilibrium criticality and dissipative criticality is all the more remarkable since, as we shall see, the critical state displays a peculiar, non-thermal scaling property of the entropy per emitter, as discussed in Sec.~\ref{s.entropy}. 
Further discussion of the correspondence between the SR transition and thermal U(1) transitions is offered in Secs.~\ref{s.mean_field_exponent}, \ref{s.g2} and \ref{s.critical_slowing_down}.

\subsection{Stability analysis and dynamical critical exponent}
\label{s.stability_TCE2}

The equations of motion Eq.~\eqref{e.dJzdt}, \eqref{e.dJpJmdt}, along with the TCE2 approximation Eq.\eqref{e.TCE2}, become a closed set of two coupled nonlinear differential equations that we can view as those of  a dynamical system. 
Introducing the variables $x_1 = \langle J^z \rangle$ and $x_2 = \langle J^+ J^- \rangle$ satisfying the equations $\dot{x}_i = f_i(x_1,x_2)$, we can build the Jacobian matrix ${\cal J}_{ij} = \frac{\partial f_i}{\partial x_j}$: 
\begin{equation}
    \mathcal{J} = 
    \begin{pmatrix}
        -\gamma_p - \gamma_d & ~ - \Gamma_N \\
        \gamma_\phi + 2 \Gamma_N \langle J^+ J^- \rangle  & ~ - \left( \gamma_p + \gamma_d + \gamma_\phi \right) + 2 \Gamma_N \langle J^z \rangle
    \end{pmatrix}~.
\end{equation}

For simplicity, we focus here on the case $\gamma_\phi = \gamma_d = 0$. The general case will be discussed in Sec.\ref{s.critical_slowing_down}. The eigenvalues of $\mathcal{J}$ take the form:
\begin{equation}
\label{e.lambda}
    \lambda_\pm = \left( \Gamma_N \langle J^z \rangle - \gamma_p \right) \pm \Gamma_N \sqrt{ \langle J^z \rangle^2 - 2 \langle J^+ J^- \rangle }~.
\end{equation}

The eigenvalues of the Jacobian matrix  probe the stability of the steady-state solution and they provide the relaxation time of the system to its stable steady state \cite{hannukainen_dissipation-driven_2018}.

At the critical point, using the solution of Eq.~\eqref{e.dJzss}, we obtain: $\lambda_+ \approx - \frac{2 \sqrt{2}}{\sqrt{N}}$ as the least negative eigenvalue, related to the direction of slowest relaxation. The related relaxation time is then $- 1 / \lambda_+$, which diverges as a power law $N^{1/2} = N^{z / d_c}$. This defines the dynamical critical exponent $z=2$, which is typical of incoherent relaxation \cite{HohenbergH1977}. As we shall further see in Sec.~\ref{s.critical_slowing_down}, based on the exact numerical solution of the problem, this exponent appears to be universal for the steady-state super-radiant transition, and it characterizes the so-called critical slowing down.

\section{Numerical results for the steady state} 
\label{s.numerical_results}

Moving beyond the cumulant truncation schemes described in the previous sections, the steady state can be reconstructed exactly for sizable systems owing to the permutational invariance of the problem, as discussed in Sec.~\ref{s.PI}.  In particular, the exact solution allows one to go beyond the observables accessible to the cumulant truncation scheme, and monitor the entropy of the steady state, its quantum correlation properties, as well as higher-order fluctuation properties related to the statistics of the emitted photons.

\subsection{Order parameter, entropy and quantum correlations across the transition} 
\label{s.qcorr}

\begin{figure}[ht!]
\begin{center}
\includegraphics[width=.9\textwidth]{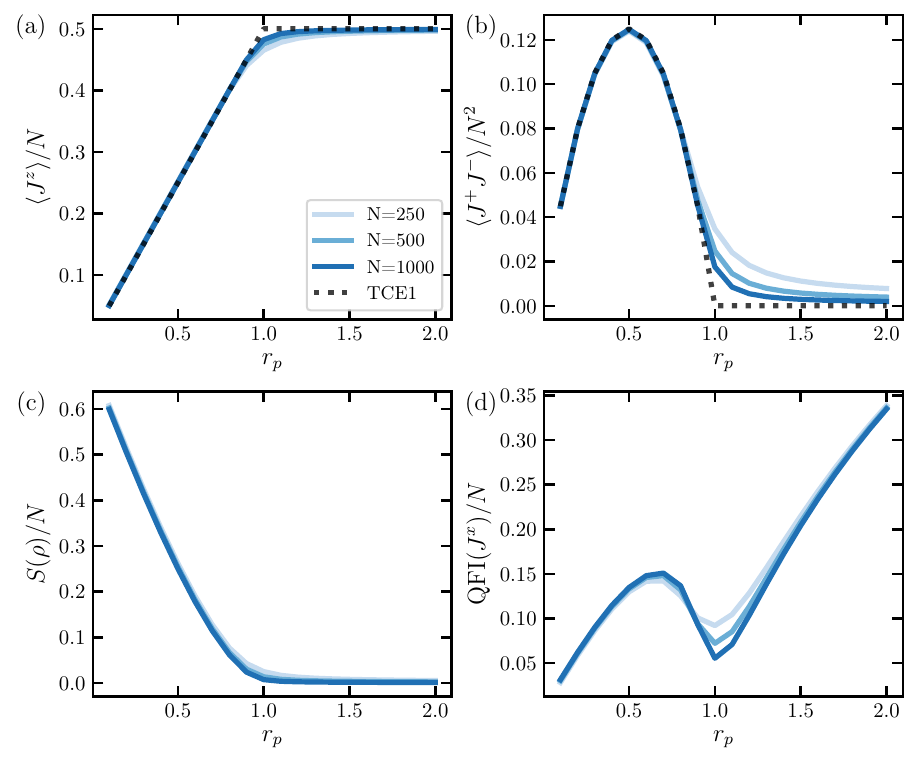}
\caption{Steady-state phase transition from super-radiance to normal emission upon increasing the pumping rate $r_p$:  (a) $\langle J^z \rangle/N$ related to the average density of excited emitters; (b) integrated, normalized dipole-dipole correlations $\langle J^+ J^- \rangle/N^2$, (c) von-Neumann entropy per emitter $S(\rho)/N = -{\rm Tr}(\rho \ln \rho)/N$ and (d) quantum Fisher information density for the operator $J^x$. All solid lines correspond to exact results. The black dashed lines correspond to TCE1 predictions, Eqs.~\eqref{e.JzTCE1} and \eqref{e.J+J-TCE1}, and are independent of N. }
\label{f.steady_state_phase_transition}
\end{center}
\end{figure}

We first consider the driven-dissipative phase transition without dephasing or individual decay, and compare the exact solution obtained for the steady state with the analytical results obtained from the cumulant approximation Eq.~\eqref{e.JzTCE1}.  In Figs.~ \ref{f.steady_state_phase_transition}(a-b), we show that TCE1 predictions Eq.~\eqref{e.J+J-TCE1}, are approached by the exact results when $N\to \infty$. We identify the super-radiant, ordered phase $0<r_p<1$ by the presence of long-range correlations $\langle J^+ J^- \rangle = O(N^2)$ and the normal, disordered phase $r_p>1$ by short-range correlations $\langle J^+ J^- \rangle / N^2 \to 0$, while $\langle J^z \rangle $ grows linearly with the pump rate $r_p$ up to the population inversion $\langle J^z \rangle = \frac{N}{2}$ in the normal phase.

A first striking difference between the driven-dissipative phase transition and a thermal phase transition, not captured by cumulant truncation schemes, lies in the entropy of the system. Indeed, the von Neumann entropy $S(\rho) = -{\rm Tr}(\rho \ln \rho)$ is found to be extensive in the ordered, super-radiant phase, but \emph{not} in the disordered, normal phase, as shown in Fig.~\ref{f.steady_state_phase_transition} (c). At thermal phase transitions, on the other hand, entropy is extensive on both sides of the transition, and in fact larger in the disordered phase than in the ordered one. The ``inverted" entropy structure at the super-radiant transition can be intuitively  understood by density-of-states considerations. The number of states $D(J,M)$ (Eq.~\eqref{e.Djm}) decreases when $J$ increases from $0$ to $N/2$. In the super-radiant phase, the steady state admixes pure states with $J < N/2$, whose number is far larger than states with next-to-extremal magnetization, \emph{i.e.}, with  $J \lesssim N/2$ and  $J^z \lesssim N/2 $, which dominate in the normal phase at large pumping strength. Hence the marked difference in entropy. Further discussion on the critical behavior of the entropy will be offered in Sec.~\ref{s.entropy}.     

 Another aspect that can be accessed via the exact solution is the quantum correlation properties, as captured by the quantum Fisher information (QFI) related to a collective spin component. ${\rm QFI}(J^\mu)$ captures the quantum contribution to the fluctuations of the collective spin component $J^\mu$ in an arbitrary mixed state (for pure states $\rho = |\psi \rangle \langle \psi |$, the QFI captures all the fluctuations, namely ${\rm QFI}(J^\mu) = 4 {\rm Var}(J^\mu)$). In particular, for the collective spin components $J^\mu$ composed of $N$ operators of spins $S=1/2$, we have that  ${\rm QFI}(J^\mu) \leq N^2$;  ${\rm QFI}(J^\mu) > N$ signals the presence of entanglement between the spins, with the QFI density  ${\rm QFI}(J^\mu)/N$ reflecting directly the depth of multipartite entanglement \cite{pezze_quantum_2018}. The states produced by the dissipative evolution are statistical mixtures of $|JM\lambda\rangle$, and therefore there are not quantum fluctuations of the $J^z$ component of the collective spin. But each of these states can be highly entangled (e.g. the state $|J=N/2, M=0\rangle$ features an entanglement depth of $O(N)$, as signaled by ${\rm QFI}(J^{x(y)})/N$). Nonetheless, in spite of the fact that in the super-radiant phase ${\rm Var}(J^{x(y)}) \sim \langle J^+ J^- \rangle \sim O(N^2)$, the quantum contribution to the fluctuations of $J^{x(y)}$ turns out to be very small, and ${\rm QFI}(J^{x(y)})$ remains systematically below $N$ throughout the phase diagram of the system, as seen in Fig.~\ref{f.steady_state_phase_transition}(d). This behavior is a direct reflection of the mixed nature of the super-radiant state, in which emitters entangle strongly with the environment and therefore are not entangled with one another.

\subsection{Robustness of super-radiant scaling to dephasing and individual decay}
\label{s.robustness_to_dephasing}

\begin{figure}[ht!]
    \begin{center}
    \includegraphics[width=\textwidth]{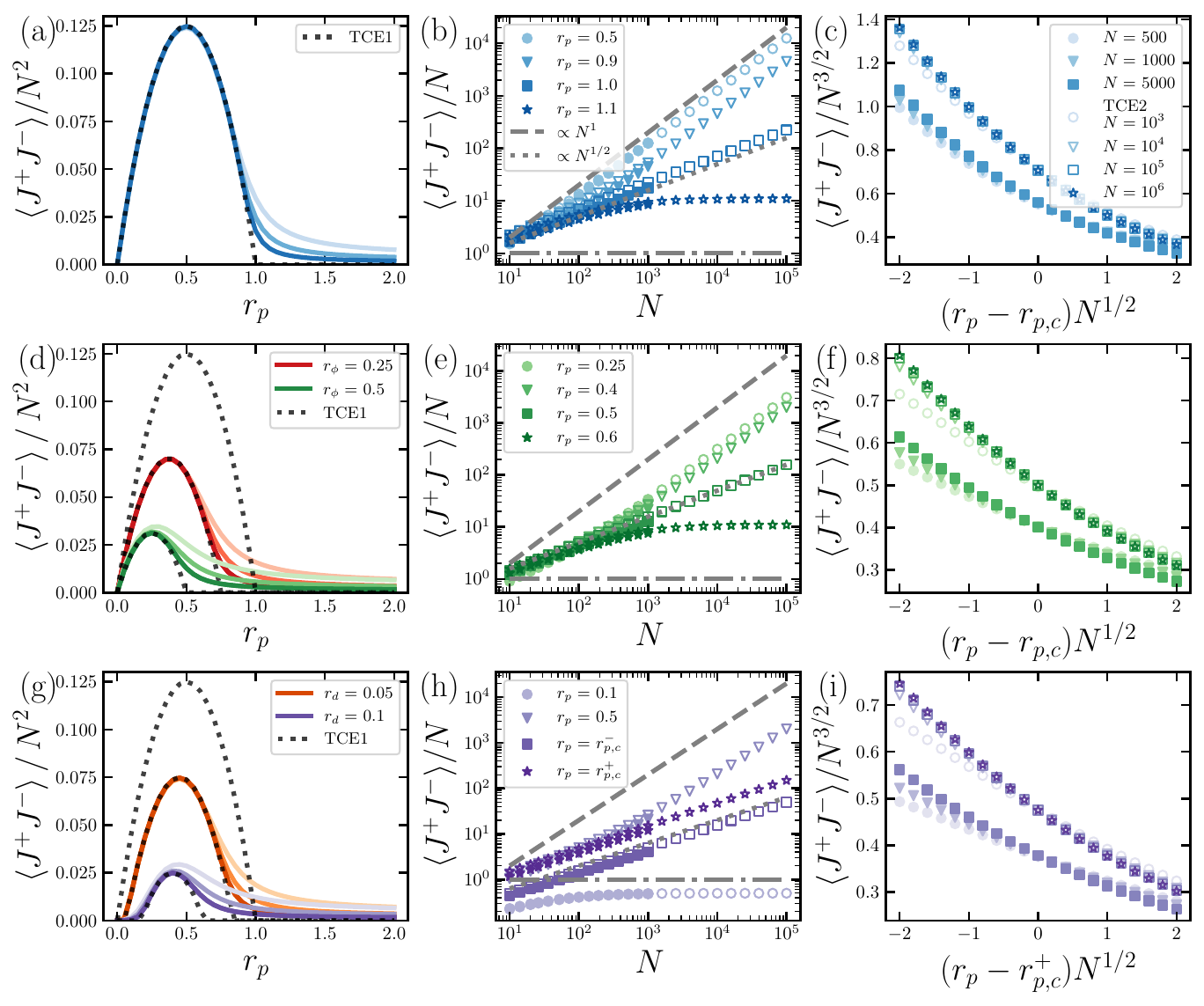}
    \caption{Exact numerical results, as well as TCE results,  for the steady state correlations across the super-radiant to normal phase transition. Panels (a,d,g) show the exact integrated correlations for three sizes ($N = 250, 500, 1000$, from lighter to darker), and for (a) $r_\phi = 0, r_d = 0$, (d) $r_\phi \neq 0, r_d = 0$, (g) $r_\phi = 0, r_d \neq 0$. Black dotted lines are TCE1 predictions, Eqs.~\eqref{e.JzTCE1} and \eqref{e.JplusJmoins_TCE1}. Panels (b,e,h) show the scaling of the integrated correlations per emitter $\langle J^+ J^- \rangle / N$ for different values $r_p$ on both sides of the phase transition and at the critical point for (b) $r_\phi = 0, r_d = 0$ (e) $r_\phi = 0.5, r_d = 0$ (h) $r_\phi = 0, r_d = 0.1$. Filled symbols are exact numerical results, whereas empty symbols are TCE2 results. Dashed and dot-dashed lines indicate scalings as $\sim N$ and $\sim N^0$, respectively. Panels (c,f,i) show the collapse of the exact results (filled symbols) and TCE2 results (open symbols) according to the scaling function defined by Eq.~\eqref{e.collapse} using mean-field universal exponents for (c) $r_\phi = 0, r_d = 0$ (f) $r_\phi = 0.5, r_d = 0$ (i) $r_\phi = 0, r_d = 0.1$.}
    \label{f.robustness}
    \end{center}
\end{figure}

In this section, we study the super-radiant phase transition in the presence of a finite dephasing rate $r_\phi \neq 0$ and a finite individual decay rate $r_d \neq 0$.

First, for $r_\phi \neq 0$, as described in Sec.~\ref{s.SR_TCE1}, one expects the phase transition to be of the same kind as the $r_\phi = 0$ case but with a critical point shifted from $r_{p,c} = 1$ to $r_{p,c} (r_\phi) = 1 - r_\phi $ (see Eq.~\ref{e.J+J-TCE1}), so that the super-radiant phase exists only for  $r_\phi < 1$. In Fig.~\ref{f.robustness}(d), we show the exact solution for the steady state correlation $\langle J^+ J^- \rangle / N^2$ through the phase transition for a few system sizes, so as to distinguish the super-radiant phase from the normal phase by the order parameter going to zero. We observe that the exact results are consistent with the predictions of the TCE1 approximation. This demonstrates that the super-radiant phase is robust to the presence of dephasing up to a critical dephasing rate $r_\phi = 1 $, beyond which the critical point drops to zero and the super-radiant phase no longer exists. The main difference with the ideal case of $r_{\phi}=0$ is the overall magnitude of the order parameter in the super-radiant phase, as shown by contrasting Fig.~\ref{f.robustness}(d) with Fig.~\ref{f.robustness}(a). 




The effect of individual decay $r_d$ is rather similar to that of dephasing in the super-radiant phase: it competes with collective decay and reduces the size of the super-radiant phase. In Sec.~\ref{s.SR_TCE1}, we obtained two solutions for the critical pumping rate given by Eq.~(\ref{e.critical_TCE1}). The first solution $r_{p,c}^-$ represents the competition between the individual pump and decay, since the pump must create a finite excitation density in the steady state to establish a macroscopic correlation between the emitters. The second solution $r_{p,c}^+$ corresponds to the transition from super-radiant to the normal phase close to full population inversion, as discussed before. In Fig.~\ref{f.robustness}(g), we show the exact steady state correlations of the system, expressed in terms of $\langle J^+ J^- \rangle / N^2$. We obtain good agreement between exact results and the TCE1 approximation when $N\gg 1$, and show that there are indeed two critical points. As an example, for $ r_d = 0.1 $, we obtain $r_{p,c}^-  \approx 0.176$ and $r_{p,c}^+ \approx 0.624$ (see the purple lines in Fig.~\ref{f.robustness}(g)).

\subsection{Critical scaling}
\label{s.critical_scaling}
In Fig.\ref{f.robustness}(b,e,h), we compare the predictions for the scaling of $\langle J^+ J^-\rangle$ obtained with the TCE2 approximation, and already discussed in Sec.~\ref{s.SR_TCE1}, and the exact numerical solution up to $N= 10^3$ emitters. We observe good agreement between the exact results and the TCE2 results for all regimes of scaling and for all strengths of couplings $r_\phi$ and $r_d$ to a local environment. In particular, we obtain the same critical point and critical scaling law with both methods. 

Interestingly, the exact results close to the critical point show a significant crossover between a scaling behavior similar to the critical one, and a behavior intermediate between critical and ordered (in the putative super-radiant phase) or between critical and disordered (in the putative normal phase). On the other hand, the TCE2 results can reach to arbitrary large system sizes, at which only three scaling behaviors are evident: the super-radiant scaling $\langle J^+ J^- \rangle \propto N^2$; the normal one $\langle J^+ J^- \rangle \propto N$; and critical scaling exactly at the critical point, $\langle J^+ J^- \rangle \propto N^{3/2}$. Further consequences of this crossover in scaling will be examined in the next subsection.

\subsection{Universal scaling functions}
\label{s.mean_field_exponent}

Given that the critical scaling at the super-radiant-to-normal transition appears to conform to the one expected at thermal U(1) phase transitions above the upper critical dimension, we push further the scaling analysis of the transition by probing the appearance of scaling functions describing the behavior around the critical point. 

Universal scaling behavior would predict that, in the vicinity of the critical point, $\langle J^+ J^- \rangle$ takes the following form
\begin{equation}
   \langle J^+ J^- \rangle  = N^{2 - 2\beta / \nu d_c} F \left( |g| N^{1/ \nu d_c} \right)
    \label{e.collapse}
\end{equation}
where $\beta = \nu = 1/2$, $d_c = 4$, and $g = r_p - r_{p,c}$ is the distance to the critical point \cite{Botetetal1982}.  

In Fig~\ref{f.robustness}(c), we show that the above scaling behavior, expected at thermal equilibrium, is well satisfied by our exact results for the dissipative transition with $r_\phi = r_d = 0$.  Previous numerical works have already observed the fact that dissipative phase transitions can exhibit a similar scaling behavior as thermal ones (see Refs.~\cite{Rotaetal2019,Verstraelenetal2023} for some relevant examples). Yet here we have the advantage of testing such scaling behavior using exact results. 
This collapse further demonstrates that the critical exponents of this phase transition are compatible with those of the mean-field universality class. 
As we shall discuss below, scaling behavior close to the critical point, complying with the existence of universal scaling functions, is observed for other quantities beyond $\langle J^+ J^-\rangle$. 

However, for the system sizes accessible to the exact solution, the scaling collapse of $\langle J^+ J^- \rangle$ is less convincing in the presence of finite dephasing or individual decay, as shown in Fig.\ref{f.robustness} (f,i). We interpret this observation to strong corrections to scaling, which are seen in the crossover behavior close to the transition, as discussed in Sec.~\ref{s.critical_scaling}. On the other hand, TCE2 results can reach much larger system sizes, and we observe good collapse for them -- modulo the fact that the scaling function appears to have changed under the TCE2 approximation, compared to the exact result. We conclude that the  super-radiant-to-normal transition in the steady state exhibits scaling behavior akin to a thermal phase transition for all its realizations considered in this work.

\subsection{Scaling laws of the entropy}
\label{s.entropy}

\begin{figure}[ht!]
\begin{center}
\includegraphics[width=\textwidth]{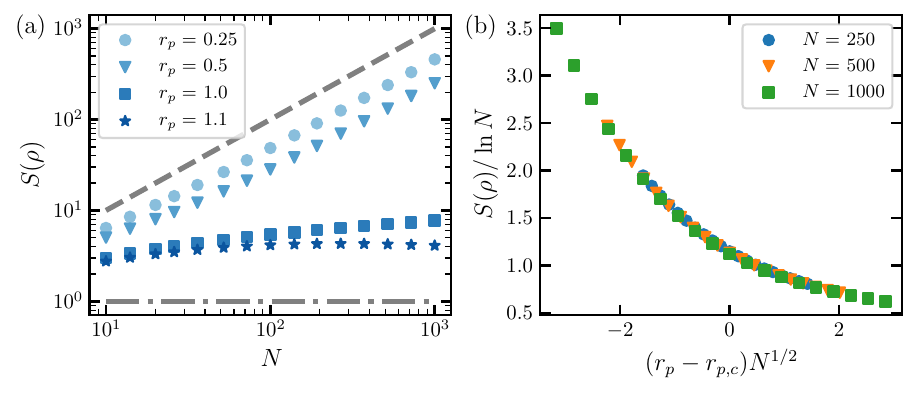}
\caption{(a) Scaling of the von-Neumann entropy $S(\rho) = - {\rm Tr} \left( \rho \ln{\rho} \right) $ of the steady state for a few values of the pumping rate $r_p$ (and $r_\phi = r_d = 0$), showing three distinct behaviours: $O(N)$ in the super-radiant phase ($r_p < 1$); $O(\ln  N)$ at the critical point ($r_p = 1$); and $O(1)$ in the normal phase, $r_p > 1 $ illustrated by a saturation plateau. (b) Collapse of the entropy around the critical point according to the scaling form of Eq.~\eqref{e.collapse_entropy}. }
\label{f.scaling_entropy_pump}
\end{center}
\end{figure}

Unlike thermal phase transitions, the super-radiant to normal phase transition is also characterized by \emph{anomalous} critical scaling of the entropy. 
The steady state super-radiant phase is generally characterized by a macroscopic entropy $S(\rho) = O(N)$ -- being associated with very large energy exchanges (and therefore entanglement) between the system and its environment. On the other hand, in the absence of local environments ($r_\phi = r_d = 0$) the normal phase is associated with a \emph{microscopic} entropy $S(\rho) = O(1)$. Detailed scaling results are shown in Fig.~\ref{f.scaling_entropy_pump}. At the critical point, the observed scaling of the entropy is consistent with a logarithmic size dependence,  $S(\rho) \approx s_0 \log N$. The scaling form for the entropy close to the critical point
\begin{equation}
S(\rho) = \log N ~F_S\left ( |g| N^{1/ \nu d_c} \right)
\label{e.collapse_entropy}
\end{equation}
leads to a very good collapse of the data for various sizes, as shown in Fig.~\ref{f.scaling_entropy_pump}(b).

\begin{figure}[ht!]
\begin{center}
\includegraphics[width=\textwidth]{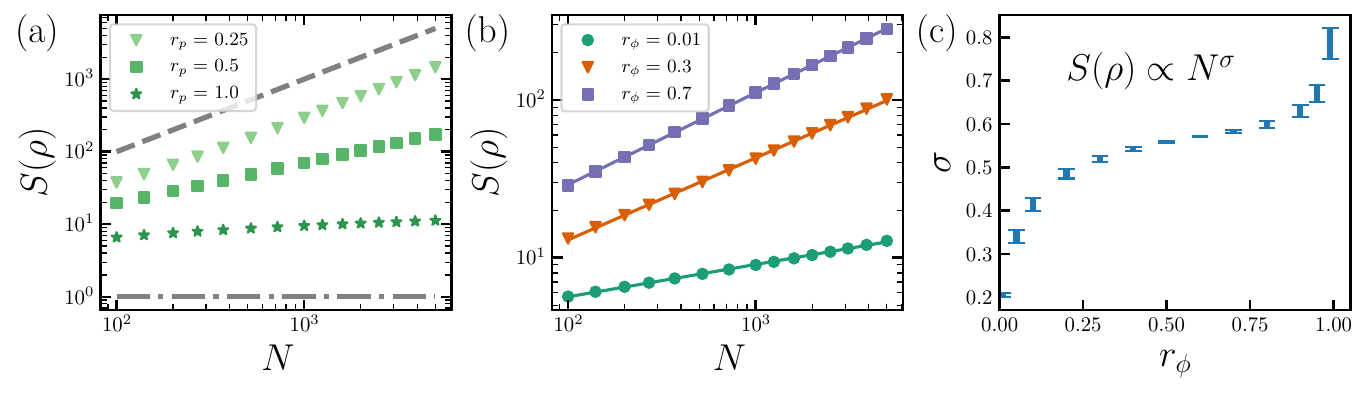}
\caption{(a) Scaling law of the entropy $S(\rho) = - {\rm Tr} \left( \rho \ln{\rho} \right)$ at fixed $r_\phi = 0.5$ for a few values of $r_p$. The dashed and dash-dotted lines indicate the scaling as $O(N)$ and $O(1)$; (b) scaling law of the entropy at the critical point $r_{p,c} = 1 - r_\phi$ for a few values of $r_\phi$. Lines correspond to a power law fit $S(\rho) \sim N^{\sigma}$; (c) exponent of a power law fit of the entropy at the critical point versus $r_\phi$. The error bars correspond to the confidence of the fit at $3 \sigma$. The first point is $r_\phi = 0.01$ and the exponent $\sigma$ vanishes when $r_\phi \to 0$.} 
\label{f.entropy_phi}
\end{center}
\end{figure}

Adding dephasing ($r_\phi \neq 0$), we observe that the scaling laws of the entropy on both sides of the transition remain unchanged, but the critical scaling of the entropy is altered, and it appears to be consistent with a power law dependence on the size,  $S(\rho) \sim N^{\sigma}$ with a non-universal exponent $\sigma$,  continuously varying with $r_\phi$ (and vanishing as $r_\phi \to 0$, such that one recovers a logarithmic scaling). These results are summarized in Fig.~\ref{f.entropy_phi}. To be precise, the entropy seems to grow with system size even in the normal phase, although, a scaling analysis reveals that it tends to saturate at large $N$. 

On the contrary, the case of a finite individual decay, $r_d \neq 0$ is fundamentally different, as the entropy remains extensive in both phases and at the phase transition. The coupling of each emitter to an individual decay channel introduces a finite entropy per emitter, and thus a macroscopic entropy throughout the phase diagram.

\subsection{Photon counting statistics and its universal critical behavior}
\label{s.g2}
\subsubsection{Dissipative transition}

An essential trait of open quantum systems is that they can be characterized not only by their internal behavior, but also by the nature of the signals they emit into the environment. 
In the case of light-emitter ensembles, a very relevant aspect for experiments is the photon statistics of the emitted light \cite{bach_emergence_2024, ferioli_emergence_2024}. A characteristic feature of this statistics is its second-order cumulant, captured by the intensity correlations at zero time delay, commonly indicated as $g^{(2)}(0)$, which for a single bosonic mode reads as $g^{(2)}(0)  = \langle a^\dagger a^\dagger  a a \rangle/ \langle a^\dagger a \rangle^2$, where $a, a^\dagger$ are operators destroying and creating photons. Establishing a correspondence between the operators for the collectively emitted light and the atomic operators \cite{allen_optical_1987}, $a \to J^{-}$, $g^{(2)}(0)$  takes the form \cite{Carmichael1980} 
\begin{equation}
    g^{(2)}(0) = \frac{\langle J^+ J^+ J^- J^-\rangle}{\langle J^+ J^- \rangle^2} = \frac{\sum_{ijkl} \langle S_i^+ S_j^+ S_k^- S_l^- \rangle }{ (\sum_{ij} \langle S_i^+ S_j^- \rangle )^2}
    \label{e.g2}
\end{equation}

\begin{figure}[ht!]
\begin{center}
\includegraphics[width=\textwidth]{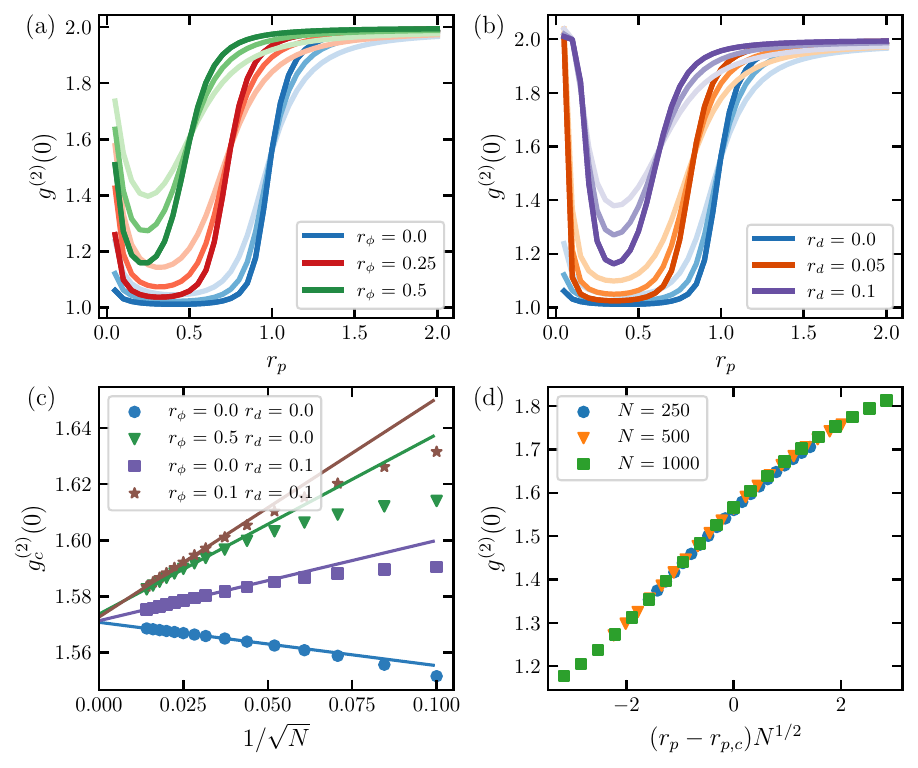}
\caption{Second-order correlation function $g^{(2)}(0)$ of the steady state across the super-radiant-to-normal phase transition (a) as a function of $r_p$ for a few values of $r_\phi$; (b) as a function of $r_p$ for a few values of $r_d$. The color gradient corresponds to increasing system sizes $N = 250, 500, 1000$; (c) $g^{(2)}(0)$ at the critical point for different parameter sets $(r_\phi, r_d)$, plotted versus $1/\sqrt{N}$. All curves converges to the universal value $U_{4,c} = \pi / 2 \approx 1.57$ in the limit $N \to \infty$, with a power-law scaling $N^{-1/2}$ according to Eq.~\eqref{e.U4}; (d)  collapse of $g^{(2)}(0)$  around the critical point for $r_\phi = r_d = 0$ according to the universal scaling function $ g^{(2)}(0) = F_g\left ( |r_p - r_{p,c}| N^{1/ \nu d_c} \right)$.  }
\label{f.steady_state_g2}
\end{center}
\end{figure}

In Fig.~\ref{f.steady_state_g2} we present exact numerical results, showing that $g^{(2)}(0)$ appears to decrease towards 1 as the system size increases in all super-radiant phases, although finite-size effects are significant, and they are amplified by the presence of local environments (dephasing and individual decay). The physical picture that emerges is therefore that of a super-radiant phase emitting light with a similar statistics to that of coherent, laser light, which has $g^{(2)}(0)=1$. On the other hand, the normal phase for strong pumping (as well as for weak pumping in the presence of individual decay) emits radiation with a similar statistics to that of thermal, incoherent light, featuring $g^{(2)}(0) \approx 2$. 
Interestingly, when transitioning from $g^{(2)}(0)\approx 1$ to  $g^{(2)}(0)\approx 2$, curves for different system sizes appear to cross close to the critical point, revealing a universal,  critical value that will be discussed further in the following section. 
Finally, we notice that the correct behavior of $g^{(2)}(0)$ in the various phases is not easily captured by truncated cumulant expansions, because of the important assumptions made by these approaches on the statistics of the spin fluctuations. We further discuss this point in App.~\ref{a.g2}.

\subsubsection{Universal Binder cumulant and critical $g^{(2)}(0)$} 

In previous sections, we have seen that the super-radiant transition exhibits universal scaling akin to a thermal $U(1)$ mean-field transition. The latter transition is also characterized by a universal, size-independent value of the so-called Binder cumulant \cite{binder_finite_1981}, which takes the form $U_4 = \langle |{\bm M}|^4\rangle / \langle |{\bm M}|^2 \rangle^2$, with ${\bm M}$ a vectorial order parameter of the transition. For a thermal phase transition above the upper critical dimension of a system with a $n$-component order parameter ${\bm M}$, the Binder cumulant takes the universal value  $U_4 (T=T_c) = U_{4,c} =  \frac{n}{4} \Gamma_E^2 \left( \frac{1}{4} n \right) / ~ \Gamma_E^2 \left( \frac{1}{4} n + \frac{1}{2} \right)$ at the transition, with $\Gamma_E(x)$ the Euler's Gamma function \cite{brezin_finite_1985}. 
In particular, a $U(1)$ transition possesses a two-component order parameter $\bm M = (J^x, J^y)$, such that the Binder cumulant at the critical point takes the value $U_{4,c}(n=2) = \pi / 2$. This universal value is a direct reflection of the dimensionality of the order parameter.

Most interestingly, one can notice that the Binder cumulant and $g^{(2)}(0)$ are actually equal up to subleading terms at the critical point (see App.~\ref{a.U4}), \emph{i.e.}, 
\begin{equation}
    U_{4,c} =  g^{(2)}_c(0)  + O \left ( \frac{1}{\sqrt{N}} \right )~.
    \label{e.U4}
\end{equation}
This implies that, when $N\gg 1$, $g^{(2)}(0)$ at a dissipative phase transition can be expected to exhibit a universal value at the critical point, in the same way as the Binder cumulant behaves at a thermal phase transition.


As shown in Fig.~\ref{f.steady_state_g2}(c), the universal value $\pi/2$ is clearly recovered numerically in the $N \to \infty$ limit for all the super-radiant transitions we considered (\emph{i.e.}, with or without local environments). The finite-size corrections to the universal value appear to consistently scale as $1/\sqrt{N}$, as suggested by Eq.~\eqref{e.U4}. Moreover, as shown in Fig.~\ref{f.steady_state_g2}(d), $g^{(2)}(0)$ follows a scaling function around the critical point, similarly to the integrated correlations and the entropy.
These results provide further evidence of the universal nature of the super-radiant transition, and its close compliance with the scaling and universality properties of a thermal transition with the same symmetry and dimensionality.  


\subsection{Critical slowing down of the dynamics} 
\label{s.critical_slowing_down}

\begin{figure}[ht!]
\begin{center}
\includegraphics[width=\textwidth]{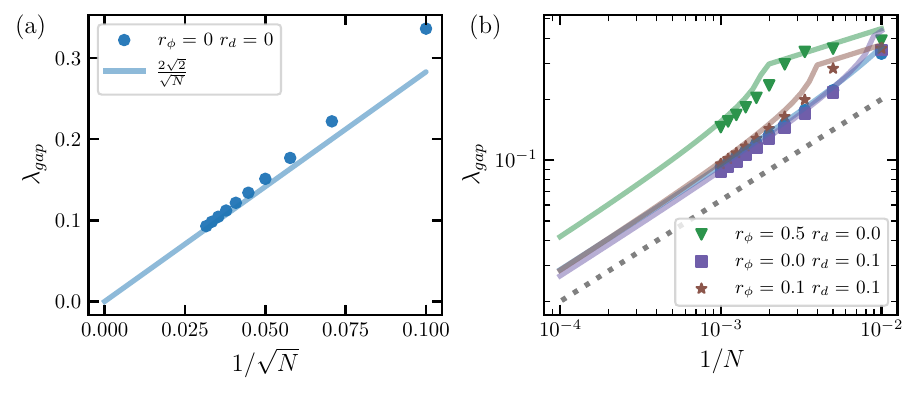}
\caption{(a) Exact results (circles) for the Liouvillian gap -- the real part of the first non-zero eigenvalue -- versus system size at the critical point for $r_\phi = r_d = 0$. The line is the TCE2 analytical prediction for large $N$ given by Eq.~\eqref{e.lambda} $- \lambda_+ = \frac{2 \sqrt{2}}{\sqrt{N}}$; (b) same quantities at the critical point for $r_\phi \neq 0$ or $r_d \neq 0$ or both. Thick lines correspond to TCE2 predictions taking into account the finite size corrections. The grey dotted line shows the power law $N^{-1/2}$. }
\label{f.gap}
\end{center}
\end{figure}

The phase transition is accompanied by a slowing down of the dynamics to reach the steady state, reminiscent of the so called critical slowing down in thermal phase transitions. By solving the exact dynamics, we obtain that the time to reach the steady state scales with system size at the critical point, with a power law $t / \Gamma_N N \propto N^{1/2} = N^{ z / d_c}$ with $z=2$ the dynamical critical exponent. This slowing down reflects a gap closing in the Liouvillian spectrum as the late time dynamics is dominated by the slowest mode. In Fig.~\ref{f.gap}, we present results for the real part of the first non-zero eigenvalue of the Liouvillian superoperator and clearly show that it goes to 0 with the critical scaling law $N^{- 1/2}$. One can note the agreement between the Liouvillian gap and the eigenvalues of the Jacobian matrix for the TCE2 solution given by Eq.~\eqref{e.lambda}, which both characterize the long time dynamics of the system at the critical point. This gap closing in also an indicator of the phase transition and its universality \cite{minganti_spectral_2018}. Fig.~\ref{f.gap}(b) shows the same behaviour for various parameters $r_\phi$ and $r_d$ at critical point. Again, TCE2 prediction is accurate to capture the universal behaviour - i.e. the gap closing at the critical point. However, TCE2 is not predictive outside of the critical point, where details of the dynamics depend on the microscopic details of the problem. Hence, it does not provide a quantitative prediction of the exact relaxation dynamics away from the critical point. More details are provided in App.\ref{a.gap}.

\section{Super-radiant burst}
\label{s.pulsed_super-radiance}

The hallmark signature of collective emission in ensembles of inverted emitters is the so-called super-radiant burst, namely a very sharp peak of emission whose intensity scales as $N^2$ \cite{dicke_coherence_1954,gross_superradiance_1982,benedict_super-radiance_2018}. It can be viewed as the transient version of the steady state super-radiance which we have discussed in the previous sections, realizing only temporarily the long-range coherence properties between emitters which are at the core of super-radiance. The qualitative features of this phenomenon have been observed in a variety of different platforms, from thermal gases \cite{skribanowitz_observation_1973,gross_observation_1976, gibbs_single-pulse_1977}, to cold atomic systems \cite{ferioli_emergence_2024,liedl_observation_2024} to solid-state emitters \cite{biliroglu_room-temperature_2022, findik_high-temperature_2021, raino_superfluorescence_2018}. In the following we aim to study the robustness of the scaling properties of the super-radiant burst to the presence of dephasing and individual emission; and the dynamical transition from super-radiant emission to collective, yet normal emission.

\subsection{Exact results without dephasing or individual decay}

We start by recalling the main properties of super-radiant bursts in the ideal case \cite{gross_superradiance_1982,benedict_super-radiance_2018}, in which the emitter ensemble is initialized in the fully excited state ($\langle J^z \rangle = N/2$), and it is then subject uniquely to collective decay, inducing long-range correlations in the system during its decay dynamics. Super-radiance is marked by the fact that, during the dissipative evolution,  $\langle J^+ J^- \rangle$ reaches values of order $O(N^2)$ around a time  $t_D \sim \tau_0 = \ln{N} /(\Gamma_N N)$ \cite{dicke_coherence_1954}. These results can be established rigorously for $N\gg 1$ from an exact solution of the dynamics \cite{benedict_super-radiance_2018}. 

\begin{figure}[ht!]
\begin{center}
\includegraphics[width=\textwidth]{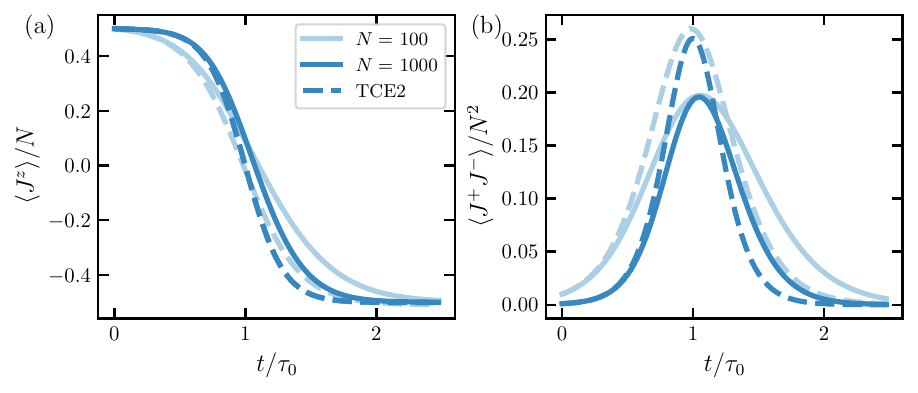}
\caption{Super-radiant dynamics starting from a state with full population inversion, comparing the exact solution (continuous lines) and the TCE2 results (dashed lines) for (a) $\langle J^z \rangle/N $ and (b) the normalized integrated correlations  $\langle J^+ J^- \rangle/N^2$. Time is normalized by the size-dependent delay time $\tau_0 = \ln{N} / \Gamma_N N $.}
\label{f.comparison_dynamics}
\end{center}
\end{figure}

In Fig.~\ref{f.comparison_dynamics}, we compare the exact dynamics (obtained numerically for any $N$) with the one obtained within the TCE2 approximation \cite{rubies-bigorda_characterizing_2023}. The approximate solution appears to overestimate the peak of correlations; yet, as we will see later in the paper, it provides the right prediction for its scaling law.

\subsection{Dynamical transition induced by dephasing}

Dephasing tends to destroy the correlations created by collective dissipation during the decay dynamics, and thus reduces the maximum correlations that the system can develop. Here, we study the scaling of the peak of emission during the dynamics, and we test whether the super-radiant scaling persists up to a critical dephasing rate. 
Similarly to what we have seen in the steady state, one can define a phase transition of the scaling properties of the transient dynamics:  $\langle J^+ J^- \rangle_{\rm max} \propto N^2$ defines a super-radiant burst; while $\langle J^+ J^- \rangle_{\rm max} \propto N$  defines the case of normal emission dynamics. Interestingly the dynamics might show a delayed emission peak even in the normal regime, due to the collective nature of the emission. Yet, if this intensity peak is only proportional to $N$, it does not correspond to super-radiance in the proper sense. It signals the dynamical buildup of correlations which nonetheless decrease with system size. Indeed, if  $\langle J^+ J^- \rangle = \sum_{ij} \langle S_i^\dagger S_j \rangle \sim N^{2-\epsilon}$ with $\epsilon > 0$, this implies that $\langle S_i^\dagger S_j \rangle_{i \neq j} \sim N^{-\epsilon}$. 

\begin{figure}[ht!]
\begin{center}
\includegraphics[width=\textwidth]{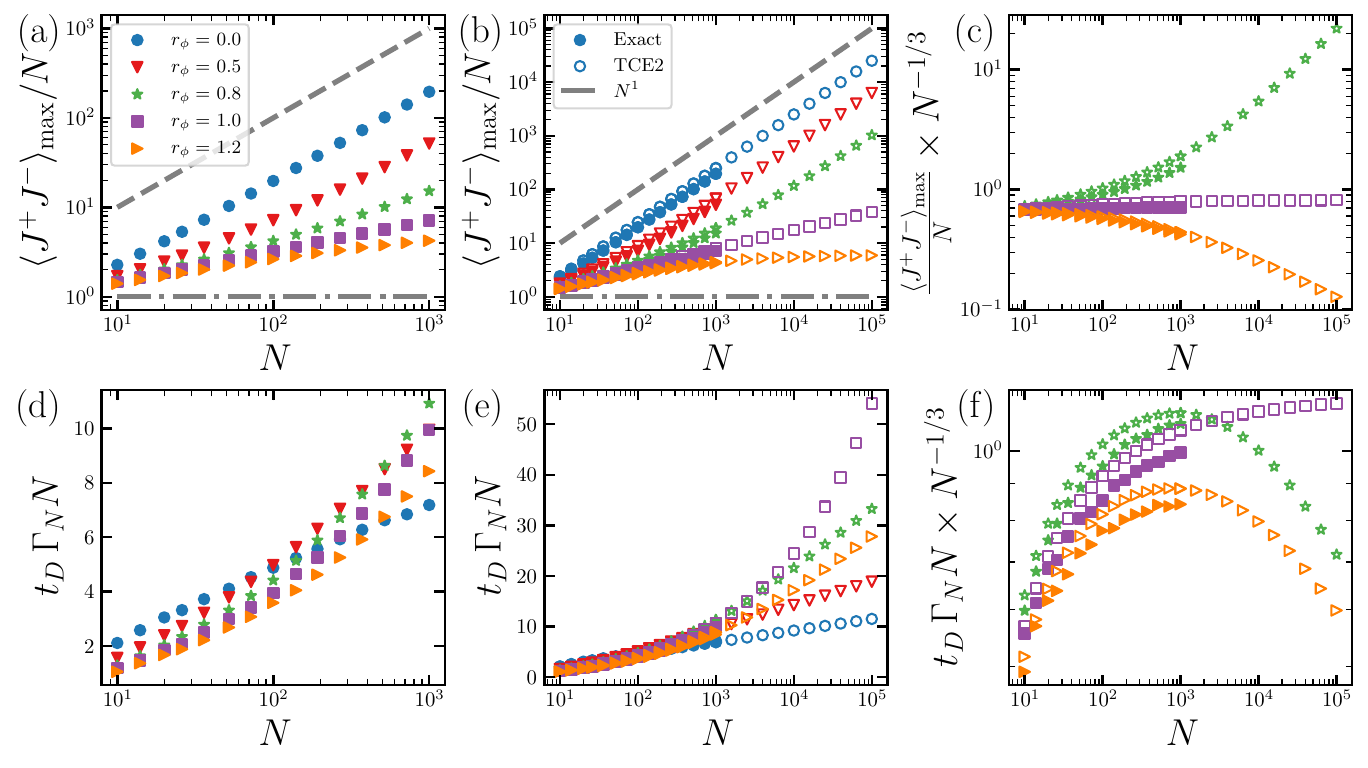}
\caption{(a,b,c) Scaling of the maximum correlations per emitter $\langle J^+ J^- \rangle_{\rm max} / N$ during the pulsed dynamics for $r_\phi \neq 0$. (a) shows the exact results (the dashed and dash-dotted lines indicating the $\sim N$ and $ \sim N^0$ scalings respectively); (b) compares them to the TCE2 ones; and (c) highlights the special scaling $\langle J^+ J^- \rangle_{\rm max} \sim N^{4/3}$ at the critical point of the dynamical transition;  (d,e,f) Time delay $t_{D}$ at which the maximum is reached. (d) presents the exact results, (e) the comparison with TCE2 results, and (f) highlights the special power-law scaling law at the critical point,  $t_D \Gamma_N N \sim N^{1/3}$.}
\label{f.dyn_scaling_phi}
\end{center}

\end{figure}

In Fig~\ref{f.dyn_scaling_phi}(a,b), we show the maximum of correlations versus the number of emitters $N$ for different values of $r_\phi$. For $r_\phi<1$ we observe emission with super-radiant scaling $O(N^2)$, whereas for $r_\phi>1$ the emission recovers normal scaling $O(N)$. In particular, we identify a specific scaling law at the critical value $r_\phi=1$, namely that $\langle J^+ J^- \rangle_{\rm max} \propto N^{\lambda}$ with $\lambda \approx 4/3$, as illustrated in Fig~\ref{f.dyn_scaling_phi}(c).  This behavior signals a dynamical phase transition between two scaling regimes, with a characteristic intermediate critical scaling. 

The critical dynamics is moreover accompanied by a significant retardation to the onset time $t_D$ of the emission peak, which connects in a non-monotonic way the size scaling of $t_D$ in the super-radiant regime and that in the normal regime. As shown in Fig.~\ref{f.dyn_scaling_phi}(d,e), throughout the super-radiant regime we observe the characteristic logarithmic scaling $t_D \sim \ln{N} / (\Gamma_N N)$ already established in the ideal case $r_{\phi} = 0$. Interestingly, this scaling is also observed in the normal phase, due to the competition between collective emission and the size dependence of the decay rate $\Gamma_N$, even if the system only develops correlations decreasing with system size. At the critical point $r_{\phi} = 1$, instead, the intensity peak is found to appear at a time  $t_D \sim N^\mu/(\Gamma_N N)$, with $\mu \approx 1/3$, as shown in Fig.~\ref{f.dyn_scaling_phi}(f).  This marks the fastest size scaling of $t_D$ with $N$ throughout the various dynamical regimes, which can be interpreted as a peculiar form of critical slowing down at a \emph{dynamical} transition.

 It is worth noticing that the sharp identification of the three distinct dynamical regimes (super-radiant, critical and normal) for various $r_{\phi}$ values requires to go to extremely large system sizes. The sizes accessible to the exact solution ($N \sim 10^3$) erroneously suggest further scaling laws, \emph{i.e.},  $\langle J^+ J^- \rangle_{\rm max} \sim N^{\lambda}$ with $\lambda$ continuously varying between 1 and 2 - see Fig.~\ref{f.dyn_scaling_phi}(a,d). In order to ascertain the correct asymptotic scaling, we make use of the TCE2 approach - see Fig.~\ref{f.dyn_scaling_phi}(b,e) - for which the system size is just a parameter, and which therefore allows for the exploration of arbitrary system sizes. Interestingly the TCE2 scheme reproduces correctly the scaling behavior observed in the exact solution at intermediate sizes; and it reveals that this scaling behavior inevitably crosses over to one of the three regimes described above ($\sim N^2$, $\sim N^{4/3}$ or $\sim N$) for very large system sizes. In App.~\ref{a.TCE1_dynamics} we show that the TCE1 approximation scheme captures the scaling law of the two phases, but it fails to capture the specific exponent at the critical point, which is only recovered to second order in the cumulant truncation. Similarly to what we have seen for the steady-state super-radiant-to-normal transition in Sec.~\ref{s.SR_TCE2}, the TCE2 approach is crucial to reconstruct the scaling properties at criticality.

\subsection{Effect of individual decay on super-radiant scaling} 

\subsubsection{Loss of super-radiant scaling at fixed density}
\label{s.loss_sr_with_nr}

\begin{figure}[ht!]
\begin{center}
\includegraphics[width=\textwidth]{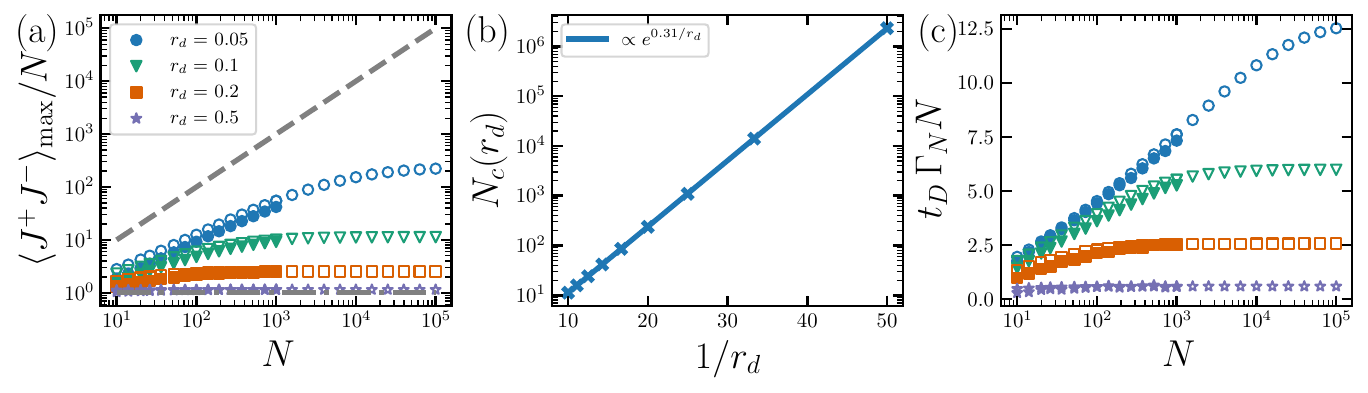}
\caption{(a) Scaling law of the maximum of correlations  $\langle J^+ J^- \rangle_{\max} / N$ during the pulsed dynamics for $r_d \neq 0$, within the fixed-density scheme. Full symbols are exact results, and open ones are TCE2 results; (b) Finite volume  of correlations, $N_c(r_d) =  \lim_{N\gg 1} \langle J^+ J^- \rangle_{\rm max}/N$, obtained with TCE2 approximation, and plotted vs $r_d^{-1}$; (c) time delay $t_{D}$ at which the maximum of correlations is reached.} 
\label{f.dyn_scaling_nr}
\end{center}
\end{figure}

We now examine the effect of individual decay on the super-radiant burst dynamics. Unlike dephasing, individual decay has a much stronger impact on the dynamics at fixed density, \emph{i.e.}, when $\Gamma_N N = \Gamma$. As shown in Fig.~\ref{f.dyn_scaling_nr}(a) a finite $r_d$ introduces the appearance of a critical size beyond which long-range correlations cannot establish anymore. While super-radiant scaling, $\langle J^+ J^- \rangle_{\rm max} \sim N^2$, might still be observed for small system sizes, when $N \gtrsim N_c(r_d)$ this scaling is found to cross over to normal scaling $\langle J^+ J^- \rangle_{\rm max} \sim N$; and concomitantly the delay time $t_D$ is also found to stop scaling -- see Fig.~\ref{f.dyn_scaling_nr}(c). This implies that individual decay introduces a characteristic finite ``volume" $N_c(r_d)$ beyond which super-radiant correlations cannot grow. Defining operationally this volume as 
$N_c(r_d) =  \lim_{N\gg 1} \langle J^+ J^- \rangle_{\rm max}/N$, we observe that it diverges very rapidly as $r_d$ vanishes, namely  $N_c(r_d) \sim \exp(R/r_d)$ with $R = 0.306 \pm 0.002$. Results are shown in Fig.~\ref{f.dyn_scaling_nr}(b). As a consequence, for sufficiently small $r_d$ there still exists a sizable range of system sizes over which super-radiant scaling can be observed. 

\subsubsection{Robustness of super-radiant scaling at fixed coupling}

\begin{figure}[ht!]
\begin{center}
\includegraphics[width=\textwidth]{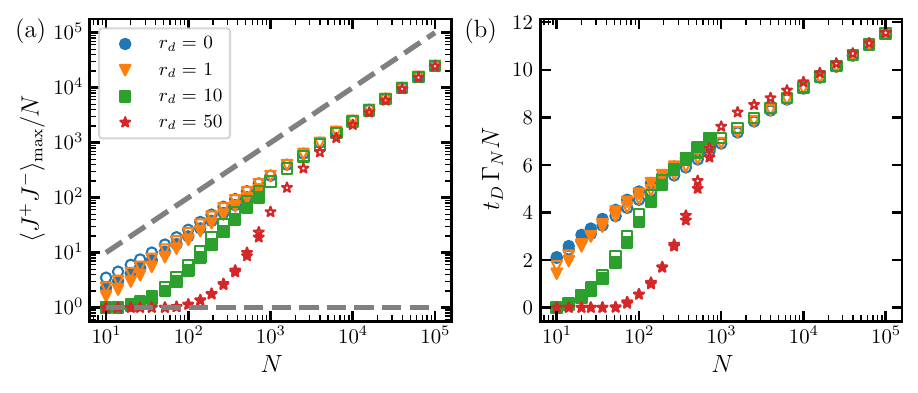}
\caption{(a) Scaling law of the maximum of correlations per emitter $\langle J^+ J^- \rangle_{\max} / N$ during the pulsed dynamics, for various values of $r_d$ within the fixed-coupling scheme $\Gamma_N = \Gamma$; (b) time $t_{D}$ at which the maximum is reached.  Full symbols are exact results, and open ones are TCE2 results. }
\label{f.dyn_scaling_nr_not_normalized}
\end{center}
\end{figure}

As seen in the previous section, the super-radiant scaling of burst dynamics does not survive the presence of individual decay in the fixed-density scenario. Yet we find it to be robust in the fixed-coupling scenario, \emph{i.e.}, when $\Gamma_N N = \Gamma N$. In this case, we observe that super-radiant scaling of the burst dynamics is recovered beyond a given threshold in the number of emitters. Intuitively this size threshold is required for collective decay, occurring at a growing rate $\Gamma N$, to outpace individual decay. The scaling behavior for selected values of individual decay rates is shown in Fig.~\ref{f.dyn_scaling_nr_not_normalized}, and it demonstrates that super-radiant scaling both for correlations and time delay is reached for all values of $r_d$ if $N$ is sufficiently large. 
Thus, within the fixed-coupling scheme super-radiant bursts are robust to any individual decay rate in the thermodynamic limit. 
However, one should keep in mind that, for a sufficiently strong individual decay, this limit may not be reached in practice, since increasing the system size within the fixed coupling scenario requires the density of the emitters to increase indefinitely.

\subsection{Photon counting statistics }

\begin{figure}[ht!]
\begin{center}
\includegraphics[width=\textwidth]{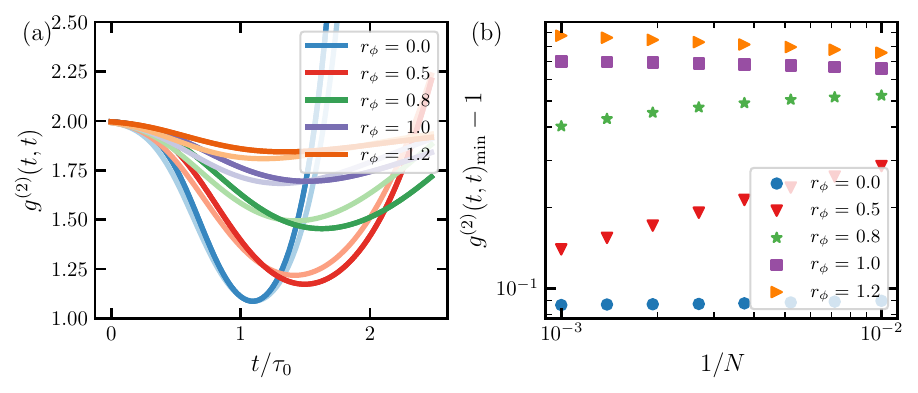}
\caption{(a) Dynamics of the second-order correlation function at equal time $g^{(2)}(t, t)$ for different values of dephasing rate $r_\phi$. The lighter and darker lines correspond to system sizes $N = 250$ and $N=500$ respectively; (b) Scaling of the minimum of $g^{(2)}(t, t)$ with system size. The data suggest a slow convergence to $1$ for $N\to \infty$ in the super-radiant phase $r_\phi<1$.}
\label{f.dyn_g2}
\end{center}
\end{figure}

We now examine the dynamics of the statistics of the photon emission, as captured by $g^{(2)}(t,t)$, \emph{i.e.}, the instantaneous  $g^{(2)}(0)$ expressing the intensity correlations at zero time delay. The definition is the same as that in Eq.~\eqref{e.g2}, but calculated on the transient state of the evolution. This quantity has been recently examined in experiments on collective light emission in atomic ensembles \cite{ferioli_emergence_2024, bach_emergence_2024}. 
In the case of a super-radiant burst, as correlations build up in the system during the dynamics, $g^{(2)}(t, t)$ is found to decrease from its initial value $g^{(2)}(0, 0) = 2$ -- which corresponds to incoherent emission --down to a minimum which is reached close to the time $t_D$ of the intensity peak. Results are shown in Fig.~\ref{f.dyn_g2}(a) for few values of dephasing rate $r_\phi$. In analogy with what we observed in the case of steady state photon statistics, in the super-radiant regime one may expect the minimum of $g^{(2)}(t, t)$  to scale down to 1, signaling coherent emission. Nonetheless we observe very strong finite-size effects, which lead to a very slow convergence of the minimum of $g^{(2)}(t, t)$ to its asymptotic limit, as shown in Fig.~\ref{f.dyn_g2}(b). Within the system sizes accessible to us, it is very difficult to conclude whether the minimum $g^{(2)}(t, t)$ converges to $1$ or to a slightly higher value. On the other hand, when $r_\phi>1$ we observe that the minimum $g^{(2)}(t, t)$ ceases to scale with system size, signaling the loss of coherent emission. 

\subsection{Evolution of purity and of quantum correlations} 
\label{s.dynamics_entropy}

\begin{figure}[ht!]
\begin{center}
\includegraphics[width=0.6\textwidth]{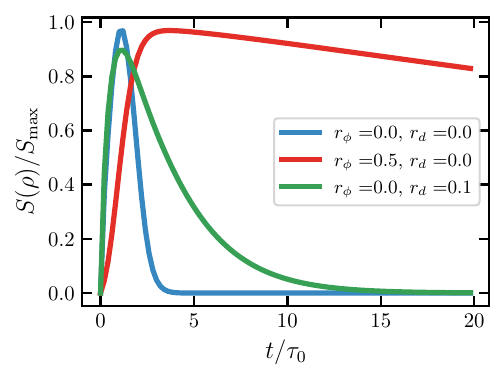}
\caption{Entropy during the decay dynamics for few values of the parameters $(r_\phi, r_d)$ and $N=100$ emitters. Results are compared to the maximal entropy, which is $S_{\rm max} = \ln{\left( N+1 \right) }$ for $r_\phi = r_d = 0$ (since only the sector of maximal spin length $J = \frac{N}{2}$ is explored); and $S_{\rm max}  = N \ln{2}$ for $r_\phi \neq 0$ or $r_d \neq 0$. }
\label{f.dynamics_purity}
\end{center}
\end{figure}

As it has been pointed out in several recent papers \cite{rosario_unraveling_2025, bassler_absence_2025, zhang_unraveling_2025}, Dicke super-radiant dynamics (without dephasing or individual decay) is associated with long-range correlations, but not with any form of entanglement; and the presence of dephasing or individual decay will certainly not promote the presence of quantum correlations. Here we offer a complementary perspective on the classical nature of the correlations in the super-radiant burst, using two different metrics (as done in Sec.~\ref{s.qcorr} for the steady state): the entropy of the emitters, as well as the quantum Fisher information of the collective-spin components. Fig.~\ref{f.dynamics_purity} shows the evolution of the von Neumann entropy of the emitters $S(\rho)$: we observe that the states close to the intensity peak of the super-radiant burst are not only the most correlated ones, but also the most entropic ones, with an entropy which comes very close to the maximum possible for $N$ emitters, i.e. $S_{\rm max} = N \log 2$ (or $S_{\rm max} =  \log (N+1)$ in the specific case $r_\phi = r_d = 0$ in which the system explores only states with maximal spin length $J = N/2$). Hence even in the dynamics we observe that super-radiance implies a very large entropy of the emitters, due to their very strong energy exchanges with the environment; this extreme entropy content is clearly at odds with the presence of quantum coherence and entanglement in the system. This further aspect is captured as well by looking at the time-dependent quantum Fisher information in App.~\ref{a.QFI}, which does not detect any form of entanglement, and which is in fact minimal around the peak of the super-radiant burst. 

\subsection{Long-time sub-radiant dynamics in the presence of dephasing} 
\label{s.subradiance}

\begin{figure}[ht!]
\begin{center}
\includegraphics[width=\textwidth]{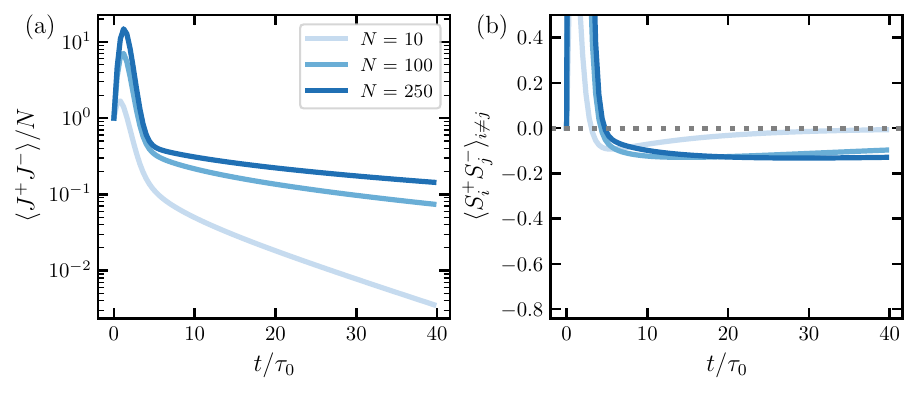}
\caption{Dynamics of correlations at long time for $r_\phi = 0.5$: (a) integrated correlations per emitter $\langle J^+ J^- \rangle / N$ and (b) correlation between two distinct emitters $\langle S_i^+ S_j^- \rangle_{i \neq j} $.}
\label{f.long_time_dynamics}
\end{center}
\end{figure}

When emitters undergo only collective emission and dephasing, we observe a two-stage dynamics. Initially, a fraction of the excitations is released in the superradiant burst; the dynamics then change qualitatively, and the remaining excitations decay with a much longer lifetime, linked to $\Gamma_N \sim O(\frac{1}{N})$. Over this second stage, the system builds up significant anti-correlations between emitters, namely $\langle S_i^+ S_j^- \rangle_{i \neq j} < 0 $.
In Fig.~\ref{f.long_time_dynamics}(a), we show the decay of the total correlations $\langle J^+ J^- \rangle$ for $r_\phi = 0.5$, highlighting two distinct lifetimes, while panel (b) focus on the emergence of anti-correlations between two emitters. Together, these results indicate that the slowdown in the dynamics originates from these anti-correlations, which manifest as destructive interference between the emission channels of the different emitters, called sub-radiance \cite{guerin_subradiance_2016}. It is worth mentioning that these anti-correlations are created only by the competition between a local environment (the dephasing $r_\phi$) and the collective dissipation, and therefore originates from a dissipative process only. They do not appear to be accompanied by any detectable form of entanglement -- \emph{e.g.}, they comply with the inequality $\sum_{\mu=x,y,z} {\rm Var}(J^\mu) \geq N/2$, satisfied by all separable states \cite{Toth2009}, but violated by entangled ones with a reduced total-spin length compared to that of \emph{e.g.}, coherent spin states.

\begin{figure}[ht!]
\begin{center}
\includegraphics[width=\textwidth]{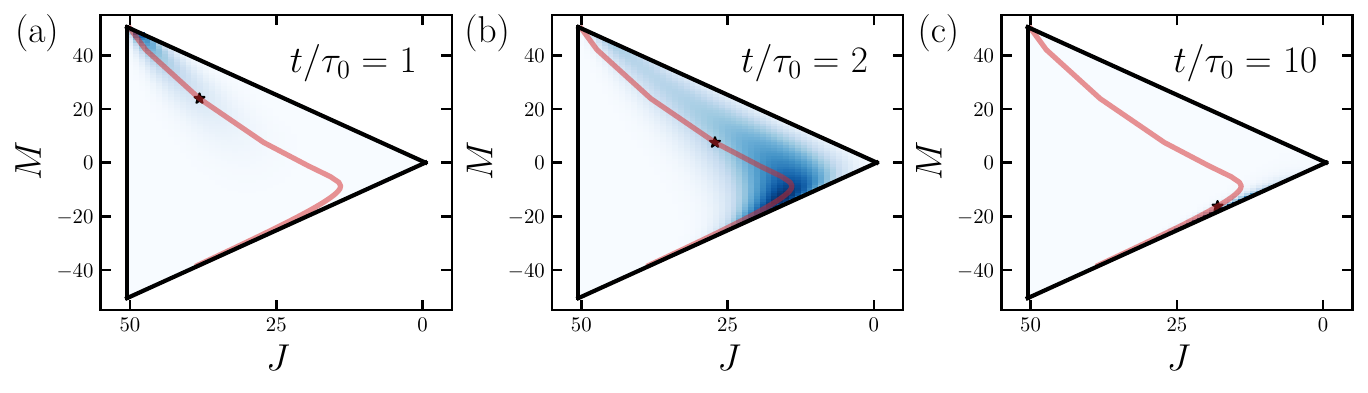}
\caption{Super-radiant dynamics for $r_\phi=0.5$, $r_d=0$, illustrated in the $(J, M)$ plane. The red line is the mean trajectory defined by $\left( \sqrt{\langle \bf{J}^2 \rangle }(t) , \langle J^z \rangle (t) \right) $. The false-color plot indicates $\rho_{J, M}$, \emph{i.e.}, the projection of the density matrix on the various $(J,M)$ sectors at time (a) $t/\tau_0 = 1$; (b) $2$, and (c) $10$. }
\label{f.dyn_triangle_phi}
\end{center}
\end{figure}

To get an intuition about the non-trivial dynamics observed with dephasing $r_\phi \neq 0$, it is useful to visualize the trajectory of this dynamics in the so-called Dicke triangle of states, \emph{i.e.}, in the plane $(J,M)$ in which each point corresponds to a group of eigenstates $|J,M,\lambda\rangle$ of ${\bm J}^2$ and $J^z$. In Fig.~\ref{f.dyn_triangle_phi}, we present the mean trajectory defined by $\left( \sqrt{\langle \bf{J}^2 \rangle }(t) , \langle J^z \rangle (t) \right) $ and few snapshots of the full density matrix $\rho = \sum_{J,M} \rho_{J, M} | J, M \rangle \langle J, M |$, at different points in time during the dynamics, which is diagonal in the illustrated basis of states $| J, M \rangle$. During the early time dynamics, the state first spreads into the bulk of the triangle. Correlations increase until the evolved state $\rho(t)$ reaches the states around $M = 0$, when the super-radiant burst attains its peak intensity. The entropy of the system is large, as described in Sec.~\ref{s.dynamics_entropy}, as the state is highly spread among different $| J, M \rangle$ states, with a significant weight on highly degenerate eigenstates of ${\bm J}^2$ and $J^z$ with small $J$. After this first burst, the state reaches the lower edge of the triangle. These states cannot decay via the collective decay channel as $J^-  | J, M=-J \rangle = 0 $. Thus, to decay, the system need to evolve to states with $M\geq -J+1$, which in turn decay with a small rate $ 2 J \Gamma_N \sim O(\frac{1}{N})$ when $J \sim O(1)$. This explains the slowdown of the dynamics, and the sub-radiant tail of the emission intensity. 

In App.~\ref{a.triangle_nr} we present a similar analysis in the presence of a finite individual decay rate and within the fixed-density scheme, although in that case we do not observe  neither super-radiance (properly defined) nor sub-radiance.

\section{Conclusions}

In this work, we have studied super-radiance, and the transition from super-radiant to normal emission, in permutationally symmetric ensembles of emitters subject to incoherent pumping. 
We have considered both the steady-state behavior under continuos pumping, as well as the transient behavior after a pulsed excitation. In each case, we have defined super-radiance in terms of the emergence of long-range correlations between the emitters; and we have investigated the robustness of these correlations to the presence of individual couplings of the emitters to the environment, such as dephasing and individual decay. Super-radiance turns out to be robust for large system sizes in the fixed-coupling scenario, in which we can increase the number of emitters without decreasing the coupling of each emitter to the light mode. On the other hand, we have demonstrated that in the fixed-density scenario, in which the decay rate of each emitter is normalized by the number of emitters, super-radiance is robust up to a critical dephasing rate or individual decay rate. 

In the steady state under incoherent pumping, the system exhibits a dissipative phase transition between a super-radiant and a normal phase.
We have shown that this transition exhibits universal scaling behavior with exponents and universal amplitudes compatible with those of a $O(n)$ model with $n=2$ and above the upper critical dimension. In particular, universal critical behavior is exhibited both by correlation properties of the emitters, as well as by the statistics of the emitted light; and the dynamical critical exponent, characterizing the slowing down of the relaxation rate to the steady state at the critical point, is also shown to be universal ($z=2$).   

In the case of a pulsed excitation, a dynamical transition appears by increasing the dephasing rate, from a super-radiant burst which exhibits long-range correlations at its peak intensity,  to a normal burst with correlations decaying with system size. The two dynamical regimes are separated by a special, critical dynamics, characterized by a novel scaling of the intensity peak of emission, as well as by a critical enhancement of the time delay to emission. Within the fixed-density scheme, individual decay suppresses super-radiant scaling of the emission intensity beyond a critical size, increasing exponentially with the inverse rate of individual decay. But super-radiant scaling is recovered instead for large sizes within the fixed-coupling scheme.  

The permutation symmetry of the system allowed us to reach unprecedented sizes within an exact study. This aspect also allowed us to benchmark the scaling predictions of the truncated cumulant scheme up to second order. The latter delivers explicit, analytical predictions for the critical scaling at the super-radiant-to-normal transition in the steady state. The combination of these approaches allowed us to unveil a rather unique example of a dissipative transition for which essentially all properties can be quantitatively established. 

Our results have important consequences for the ongoing experimental studies of super-radiance in \emph{e.g.}, solid-state emitters subject to a dephasing thermal bath, as well as to individual decay. We establish the strict conditions of robustness of super-radiance, and the main finite-size limitations to its observation. As an example, in light of our results the recent observation of burst intensities scaling as $N^{\alpha}$ with $\alpha = 1.6\div 1.7$ in super-crystals of perovskite nano-cube emitters \cite{raino_superfluorescence_2018, biliroglu_room-temperature_2022} could be interpreted as possible finite-size crossovers to proper super-radiant scaling ($\sim N^2$) in the vicinity of the transition to normal scaling. 

This work offers a new theoretical perspective on a phenomenology that is already well known, which is super-radiance of an ensemble of emitters invariant under permutation. Here, we have worked to describe superradiance as a phase of matter and to apply the tools of statistical physics to it. We have benefited from the permutational invariance symmetry of the problem to access numerically really large system sizes that allowed us to draw unequivocal conclusion about the scaling laws and to benchmark truncated cumulant expansion on this problem. This study can serve as a basis for more complex study of driven dissipative phase transition and their universality. Permutational invariance -- an essential ingredient of our work --  can be realized in experiments, both for emitters in free space as well as in cavities. Yet future directions of this work will naturally include testing whether the features of super-radiance and its transition to normal emission are robust to the loss of permutational invariance. A related question concerns the role of dimensionality of the emitter array (without permutational invariance) in the existence of super-radiance under incoherent pumping, and in the nature of its transition to normal emission.

\bigskip \emph {Note added.} While completing this manuscript, we became aware of Ref.~\cite{bassler2026scalingtheorydecoherencedicke}, which, similarly to our work, investigates the scaling properties of super-radiant bursts in the presence of dephasing and individual decay within a mean-field framework, supplemented with exact results. Wherever there is overlap, the results of Ref.~\cite{bassler2026scalingtheorydecoherencedicke} and ours agree.


\section*{Acknowledgements}
We thank B. Ab\'ecassis, B. Besga, L. Coolen, J. Houel and B. Mahler for important exchanges that sparked this project, and we thank them as well as J. Dubail for very useful discussions.  


\paragraph{Funding information}
This work is supported by Agence Nationale de la Recherche (ANR-25-CE57-
1786, project STEFAN, and ANR-22-PETQ-0004
France 2030, project QuBitAF). L.R. is supported by France 2030 project QuanTEdu-France ANR-22-CMAS-0001.

\begin{appendix}
\numberwithin{equation}{section}


\section{Refined version of the truncated cumulant expansion }
\label{a.TCEb}

In the main text, we have applied the truncated cumulant expansion (TCE) at the level of collective spin variables. This approach introduces a systematic error, since it does not fully respect the spin algebra for $S=1/2$ spins.
For example, writing $\langle J^+ J^- \rangle \approx \langle J^+ \rangle \langle J^- \rangle$ within the TCE1 scheme assumes that $\langle S_i^+ S_i^- \rangle = \langle S_i^+ \rangle \langle S_i^- \rangle $ instead of $\langle S_i^+ S_i^- \rangle = \langle S_i^z \rangle + \frac{1}{2}$.
A more refined version of this approximation consists in applying the factorization of averages of products into products of averages uniquely to commuting variables, \emph{i.e.}, for correlators between different spins $\langle S_i^\mu S_j^\nu \rangle \approx \langle S_i^\mu \rangle \langle S_j^\nu \rangle $ for $i \neq j$ and to treat exactly the terms with $i=j$. Within permutational invariance between spins, we can redefine the first-order truncation scheme as $\langle J^\mu J^\nu \rangle =: (1- \frac{1}{N}) \langle J^\mu \rangle \langle J^\nu \rangle + \frac{i}{2} \sum_\gamma \epsilon_{\mu\nu\gamma} \langle J^\gamma \rangle$.
This distinction between cumulant truncation for collective or individual spin variables is already discussed in \cite{ritsch_benchmarking_2025}. 
The difference between the cumulant truncation scheme at the level of the collective spin operators and that at the level of spin-spin correlators amounts to terms of order $O(N)$, which are generally subdominant in terms of scaling with respect to the term $\langle J^+ \rangle \langle J^-\rangle$, scaling as $O(N^2)$ in the super-radiant regime. 
Applying the TCE1 approximation only at the level of commuting variables, we get the set of equations:

\begin{eqnarray}
    \frac{d \langle J^z \rangle}{dt}  &=&  \gamma_p \left( \frac{N}{2} - J^z \right) - \gamma_r \left( \frac{N}{2} + J^z \right) - \Gamma_N \left( \left( 1 - \frac{1}{N} \right) J^+ J^- + \left( \frac{N}{2} + J^z \right) \right)  \\
    \frac{d \langle J^+ \rangle}{dt} &=& - \frac{1}{2} \left( \gamma_p + \gamma_\phi + \gamma_r \right) J^+   + \Gamma_N \left(  \left( 1 - \frac{1}{N} \right) J^+ J^z  - \frac{1}{2} J^+  \right) 
\end{eqnarray} 
The steady state solution of these equations are: 
\begin{enumerate}
\item for $r_p < r_{p,c}$:
\begin{eqnarray}
    \langle J^z \rangle &=& \frac{N}{2} \frac{N}{N-1} \left( r_p + r_\phi + r_d + \frac{1}{N} \right) \nonumber \\
    |\langle J^+ \rangle |^2 &=&   \frac{N^2}{2} \frac{N}{N-1} \left( r_p - r_d - \frac{1}{N} \right) - \frac{N}{N-1} \left( r_p + r_d + \frac{1}{N}  \right) \left( r_p + r_\phi + r_d + \frac{1}{N} \right) \nonumber
\end{eqnarray}

\item 
for $r_p > r_{p,c}$:
\begin{equation}
    \langle J^z \rangle = \frac{N}{2} \frac{r_p - r_d - \frac{1}{N}}{r_p + r_d + \frac{1}{N}} 
     ~ ~ ~ ~ ~ ~ ~ ~ ~ ~ ~ ~ ~ ~ \langle J^+ \rangle = 0 \nonumber
\end{equation}
\end{enumerate}
which gives a correction of order $O(1)$ to $\langle J^z \rangle$ and of order $O(N)$ to $|\langle J^+ \rangle |^2$ compared to Eq.\eqref{e.JzTCE1} and \eqref{e.JplusJmoins_TCE1}.
Therefore, the refined TCE1 approximation only adds sub-leading size-dependent term, that do not modify the TCE1 predictions for the asymptotic scaling behavior.  Especially, it does not recover the specific scaling $O(N^{3/2})$ observed at the critical point in the exact solution and via the TCE2 approach. 

A similar refinement of the truncation scheme can be applied to the TCE2 approach by applying the factorization of third order correlators uniquely for different sites, namely writing $\langle S_i^+ S_j^z S_l^- \rangle =: \langle S_i^+ S_l^- \rangle \langle S_j^z \rangle$ only when $i \neq j \neq l$. Then one obtains:
\begin{equation}
    \langle J^+ J^z J^- \rangle =:  \langle J^+ J^- \rangle \langle J^z \rangle - \left ( \frac{N}{4} +  \langle J^z \rangle +  \frac{\langle J^z \rangle^2}{N} \right ) + N(N-1) \left [ C^{zz} - \frac{\langle J^z\rangle^2}{N} - C^{+-} \left (1 + \frac{2\langle J^z \rangle)}{N} \right ) \right ] 
\label{e.J+JzJ-}
\end{equation}
where the corrections come from treating exactly the case $i=j=l$, and $i=j\neq l$ (plus permutations). 
Here $C^{\mu\nu} = \langle S_i^\mu S_{j\neq i}^\nu \rangle$ is an off-site correlator, which, for a permutationally invariant system, can be written as 
\begin{eqnarray}
\label{e.offsite}
C^{+-} & = & \frac{1}{N(N-1)} \left ( \langle J^+ J^- \rangle - \langle J^z \rangle - \frac{N}{2} \right )  \nonumber \\
C^{zz} & = & \frac{1}{N(N-1)} \left ( \langle (J^z)^2 \rangle - \frac{N}{4} \right ) ~.
\end{eqnarray}
Again, the difference between the cumulant truncation scheme at the level of the collective spin operators Eq.~\eqref{e.TCE2} and that at the level of spin-spin correlations, Eq.~\eqref{e.J+JzJ-}, amounts to terms of order $O(N^2)$ in Eq.~\eqref{e.J+JzJ-}, which are generally subdominant in their size scaling compared with the term $\langle J^+ J^- \rangle \langle J^z \rangle$. The latter scales as $O(N^3)$ in the super-radiant regime. This new scheme requires one to track the evolution of the $C^{zz}$ correlator, whose equation of motion -- within the same approximation scheme -- reads: 
\begin{equation}
  \dfrac{d C^{zz}}{dt} = \left( \gamma_p - \gamma_d - \Gamma_N \right) \frac{\langle J^z \rangle}{N} 
 -  2 \left(\gamma_p + \gamma_d + \Gamma_N \right) C^{zz} - \Gamma_N C^{+-} - 2 ~\frac{N-2}{N}~ \Gamma_N ~ C^{+-} \langle J^z \rangle~. 
\end{equation}
This equation must be supplemented with those for the magnetization $\langle J^z \rangle$ and the $C^{+-}$ correlator: 
\begin{eqnarray}
    \dfrac{d \langle J^z \rangle}{dt} &=&   \left( \frac{\gamma_p - \gamma_d - \Gamma_N}{2} \right)  - \left(\gamma_p + \gamma_d + \Gamma_N \right) \langle J^z \rangle -  N \Gamma_N (N-1) C^{+-} \\
    \dfrac{d C^{+-}}{dt} &=&  - \left(\gamma_p + \gamma_\phi + \gamma_d + \Gamma_N \right) C^{+-}  
    + 2\Gamma_N C^{zz} + \Gamma_N \frac{\langle J^z \rangle}{N} +  \frac{2 \Gamma_N (N-2)}{N} \langle J^z \rangle  \nonumber~.
\end{eqnarray}

Even if this refined scheme is more accurate than the one applied directly at the level of the collective spin variables (Eq.~\eqref{e.TCE2}), we use the latter one for the results shown in the main text, due to the simplicity of its mathematical expressions; and because the predictions of the dominant scaling behavior are not altered by the refinement of the approach. See also the next section fo the higher-moment correlators entering in $g^{(2)}(0)$. 

\section{Photon statistics via cumulant truncation schemes} 
\label{a.g2}


In this section we present TCE results for the second-order correlation $g^{(2)}(0)$ in the steady state.


Using the notation defined in Eq.~\ref{e.offsite}, we can rewrite $g^{(2)}(0)$ in terms of off-site correlators:

\begin{eqnarray}
    g^{(2)}(0) &=& \Bigg [ N(N-1)(N-2)(N-3) C^{++--} + 4N(N-1)(N-2) \left( C^{+z-} + \frac{1}{2} C^{+-} \right) \nonumber \\
    && +  2N(N-1) \left( C^{zz} + C^z + \frac{1}{4} \right ) \Big ]~ /~ \left(N(N-1) C^{+-} + N C^z + \frac{N}{2} \right)^2 
     \label{e.g2_all_terms}
\end{eqnarray}
where $C^\mu= \langle S_i^\mu \rangle$,  $C^{\mu\nu} = \langle S_i^\mu S_j^\nu \rangle$, $C^{\mu\nu\lambda} = \langle S_i^\mu S_j^\nu S_k^\lambda \rangle$ and  $C^{\mu\nu\lambda\rho} = \langle S_i^\mu S_j^\nu S_k^\lambda S_l^\rho \rangle$, in which correlators involve only different sites.




The asymptotic behaviour of the the second-order correlation function $g^{(2)}(0)$ on both sides of the super-radiant-to-normal transition can be obtained by analysing the dominant scaling terms within the TCE1 approximation, yet using the refined approach explained in App~\ref{a.TCEb}.
Within this approach, all terms can be written in terms of $C^\mu$ values; keeping only the dominant ones in $N$, one gets:
\begin{itemize}
    \item $C^+ \neq 0 \Rightarrow g^{(2)}(0) \approx \frac{N^4 |C^+|^4}{ \left( N^2 |C^+|^2 \right)^2} = 1$
    \item $C^+ = 0 \Rightarrow g^{(2)}(0) \approx \frac{2 N^2 \left( C^{zz} + C^z + 1/4 \right) }{ \left( N \left(C^z + 1/2 \right) \right)^2} = 2$
\end{itemize}
which correspond to the correct asymptotic limit of the super-radiant and normal phase -- see Fig.~\ref{f.steady_state_g2}.

On the contrary, if we apply TCE2 approximation to Eq.~\ref{e.g2_all_terms}, one obtains $C^{++--} \approx 2 C^{+-} C^{+-}$ and thus $g^{(2)}(0) \approx 2 $ in the super-radiant phase with $N\gg 1$ (with $C^{+-}$ finite), which is contrary to the exact result $g^{(2)}(0) \approx 1 $. This difference is due to the nature of the TCE2 approximation, which is akin to a Gaussian Ansatz for the spin statistics, fundamentally limiting the kind of higher-order fluctuations it can describe. In particular, the TCE2 approach delivers systematically $g^{(2)}(0) \to 2$ (for $N \gg 1$) in all regimes. 
In this case, TCE2 approximation describes well the mean value of product of one or two operators, but cannot describe correctly higher-order moments. 

\begin{figure}[ht!]
\begin{center}
\includegraphics[width=0.6\textwidth]{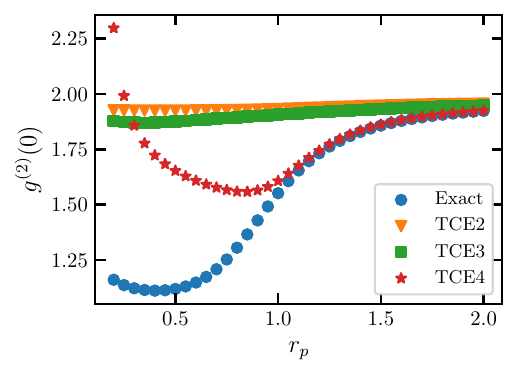}
\caption{Second-order correlation $g^{(2)}(0)$ of the steady state for $N=100$ emitters and $r_\phi = r_d = 0$. Comparison between the exact results, and different orders of truncation in the cumulant approximation. Here the truncation is done at the level of individual emitters as described in App.~\ref{a.TCEb}.}
\label{f.appendix_b}
\end{center}
\end{figure}

Following this observation, one could expect that going higher in the order of the cumulant truncation scheme, one could capture correctly higher order moments of the spin statistics. 
In Fig.\ref{f.appendix_b}, we show the predictions for $g^{(2)}(0)$ in the steady state for different orders in the truncation up to \emph{fourth} order. Here the truncation is done at the level of individual spin, as described in App.\ref{a.TCEb}. Even if higher-order TCEs give better predictions, they fails to describe correctly the super-radiant phase, for all orders considered here. $g^{(2)}(0)$ is highly sensitive to the exact statistics of the spin variables, and all TCEs make strong assumptions on the higher moments of theses statistics. This is an important consideration beyond the calculation of $g^{(2)}(0)$. The truncated cumulant expansion proves to be an excellent approximation for calculating the mean values of first and second moment of observables, but struggles to provide accurate information about the precise distribution of these observables.

\section{Link between Binder cumulant and $g^{(2)}(0)$} 
\label{a.U4}

In this section, we derive the relation between the Binder cumulant and the second-order correlation $g^{(2)}(0)$. Considering the $U(1)$ symmetry of our problem, the two-component order parameter is $\bm M = (J^x, J^y)$, such that the relation between the order parameter and the intensity is:
\begin{equation}
  |{\bm M}|^2 = (J^x)^2 + (J^{y})^2  = J^+ J^- - J^z
\end{equation}
By replacing this expression in the definition of the Binder cumulant \cite{binder_finite_1981}, we get :
\begin{equation}
    U_4 = \frac{\langle |{\bm M}|^4\rangle}{ \langle |{\bm M}|^2 \rangle^2 }   
    =  \frac{\langle J^+ J^+ J^- J^- - 4 J^+ J^- J^z + 2 J^+ J^- + J^z J^z \rangle}{\langle J^+ J^- - J^z \rangle^2} ~. 
\end{equation}
At the critical point, one expects $\langle J^+ J^- \rangle = O(N^{3/2})$ and $J^z = O(N)$ such that, keeping only the dominant terms, it simplifies to:
\begin{equation}
   U_{4,c} \approx \frac{\langle J^+ J^+ J^- J^- \rangle + O(N^{5/2})}{\langle J^+ J^- \rangle^2 + O(N^{5/2})} ~.
\end{equation}
Assuming $\langle J^+ J^+ J^- J^- \rangle \sim O(N^3)$, we obtain an identity between the Binder cumulant and $g^{(2)}(0)$ up to finite-size corrections of order $O(N^{-1/2})$ -- Eq.~\eqref{e.U4}.

\section{Stability analysis of TCE2 solution}
\label{a.gap}


\begin{figure}[ht!]
\begin{center}
\includegraphics[width=0.6\textwidth]{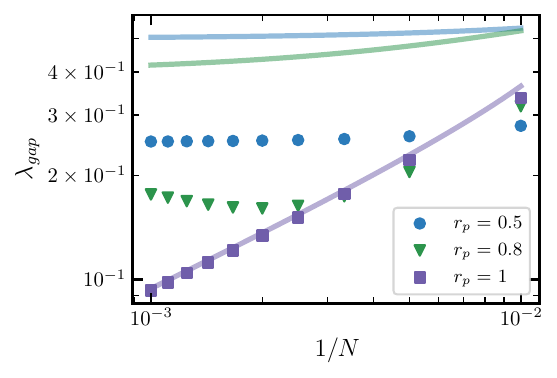}
\caption{Liouvillian gap (real part of the first non-zero eigenvalue) versus system size 
for few values of $r_p$ and $r_\phi = r_d = 0$. Lines correspond the eigenvalue of the Jacobian matrix given by the TCE2 equations, taking into account finite size corrections. $r_p = 1$ is the critical point for which the Liouvillian gap and the eigenvalue of the Jacobian matrix coincide, exhibiting the critical scaling law $N^{-1/2}$. }
\label{f.appendix_stability}
\end{center}
\end{figure}

The stability analysis of the equation of motion carried out in Sec.~\ref{s.critical_slowing_down} can also be performed away from the critical point. In the super-radiant phase, we get from Eq.~\eqref{e.lambda}, keeping only the leading order in $N$: 
\begin{equation}
    \lambda^\pm = - \frac{\gamma_p}{2} \pm \frac{1}{2} \sqrt{5 \gamma_p^2 - 4 \gamma_p \Gamma_N N}
\end{equation}
The (real) eigenvalue of the Jacobian matrix with the smallest absolute value is $\lambda^+$, and it corresponds to the slowest time scale of the dynamics. The characteristic time needed to reach the steady state is then proportional to $-1/\lambda_+$.  Fig.~\ref{f.appendix_stability} shows that  $\lambda^+$ predicted by the TCE2 approach is in good agreement with the exact Liouvillian gap at the critical point of the steady-state super-radiant-to-normal transition, at which the gap scales to zero for large sizes $N$. Yet this agreement becomes only qualitative when moving into the super-radiant phase, \emph{i.e.}, when the gap remains finite. 

\section{Mean-field solution of the pulsed dynamics}
\label{a.TCE1_dynamics}

\begin{figure}[ht!]
\begin{center}
\includegraphics[width=\textwidth]{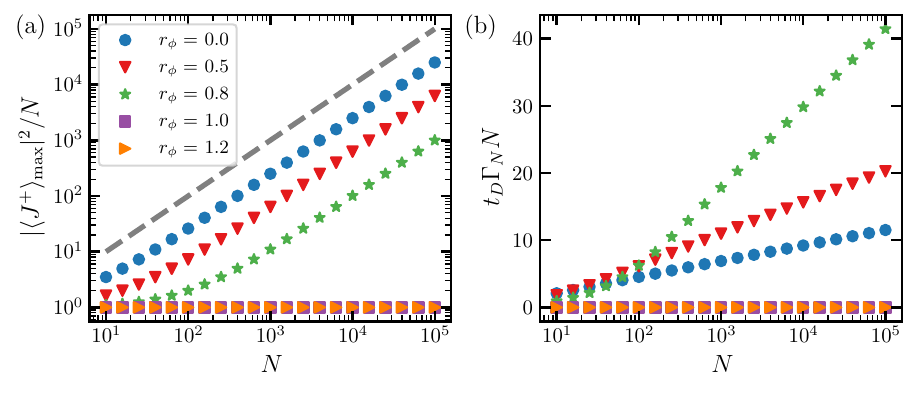}
\caption{TCE1 results for the super-radiant burst dynamics for various dephasing rates $r_\phi$: (a) scaling of the peak intensity during the decay dynamics; and (b) scaling of the time of the peak.}
\label{f.appendix_d}
\end{center}
\end{figure}

The TCE1 approximation is sufficient to capture the essential physical features of the super-radiant burst, even though it  agrees quantitatively with the exact solution only in thermodynamic limit. 
By solving the dynamics of Eq.~\eqref{e.dJzdt} and \eqref{e.dJpdt}, with $\gamma_p = 0$, starting from the fully inverted state $\langle J^z \rangle = \frac{N}{2}$ and assuming a non-zero initial value of $\langle J^+ \rangle$, the TCE1 scheme can correctly predict the scaling of the peak intensity and the time at which this peak is reached for the super-radiant phase.  Results are shown in Fig.~\ref{f.appendix_d}. 
However, it does not capture the specific scaling at the critical point, nor the residual collective effects in the normal phase. This fully justifies to go up to second order in the cumulant truncation in order to quantitatively describe the dynamical transition.

\section{Dynamics in the Dicke triangle with individual decay}
\label{a.triangle_nr}

\begin{figure}[ht!]
\begin{center}
\includegraphics[width=\textwidth]{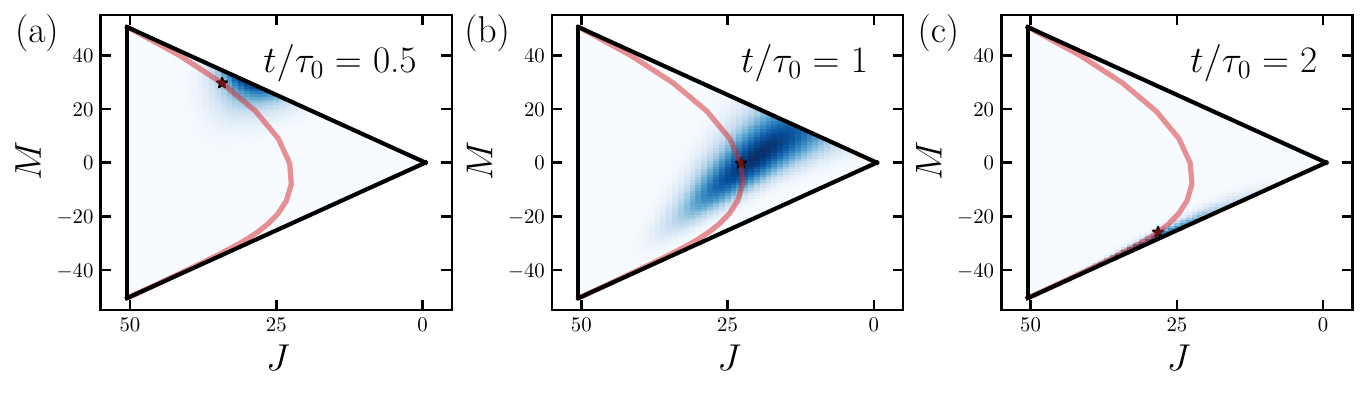}
\caption{Super-radiant dynamics for $r_\phi=0$, $r_d=0.1$, illustrated in the $(J,M)$ plane. The red line is the mean trajectory defined by $\left( \sqrt{\langle \bf{J}^2 \rangle }(t) , \langle J^z \rangle (t) \right) $. The false-color plot indicates $\rho_{J, M}$, \emph{i.e.}, the projection of the density matrix on the various $(J,M)$ sectors at time (a) $t/\tau_0 = 0.5$; (b) $1$; and (c) $2$.  }
\label{f.dyn_triangle_nr}
\end{center}
\end{figure}

In this section, we illustrate the evolution of the state of the system in the Dicke triangle during the dissipative dynamics induced  by collective emission as well as individual decay.
Fig.~\ref{f.dyn_triangle_nr} shows a few representative snapshots of the $\rho_{J,M}$ distribution in the $(J,M)$ plane, as well as the mean trajectory, similar to what we discussed in Sec.~\ref{s.subradiance} for the case of dephasing. The mean trajectory shows an apparent similarity to the latter case, yet the full distribution shows important differences, highlighting why the super-radiant burst is robust to dephasing but not to individual decay. As one can see in Fig.~\ref{f.dyn_triangle_nr}(a), in the early time dynamics the state moves away from maximal-$J$ states, yet it remains localized close to the upper edge of the triangle -- unlike what happens in the presence of dephasing, when the state enters in the bulk of the triangle, compare Fig.~\ref{f.dyn_triangle_nr}. 
As a consequence, with individual decay the emitters are prevented from reaching states with $M\approx 0$ and $J \sim O(N)$, which are the ones carrying truly long-range correlations, namely $\langle J^+ J^- \rangle \approx J(J+1) \sim O(N^2)$; these states are instead reached in the ideal case, or in the presence of dephasing below the critical threshold.  When increasing the system size in the presence of individual decay, the state remains close to the upper edge until it reaches $J$ values with $J\sim O(1)$; as a consequence, the integrated correlations do not scale quadratically with $N$ for arbitrary sizes, but only for small sizes, as shown in Sec.~\ref{s.loss_sr_with_nr}.

After the peak intensity, the state reaches the lower edge of the triangle. In contrast to the case with dephasing, states with quantum numbers $(J,-J)$ can decay directly to states with $(J+1,-J-1)$ via the individual decay. Hence the long time dynamics is an exponential decay with the exponent defined by the individual decay rate. As a consequence, sub-radiance at long times is absent unlike the case of dephasing, Sec.~\ref{s.subradiance}.

\section{Quantum Fisher Information}
\label{a.QFI}

\begin{figure}[ht!]
\begin{center}
\includegraphics[width=0.6\textwidth]{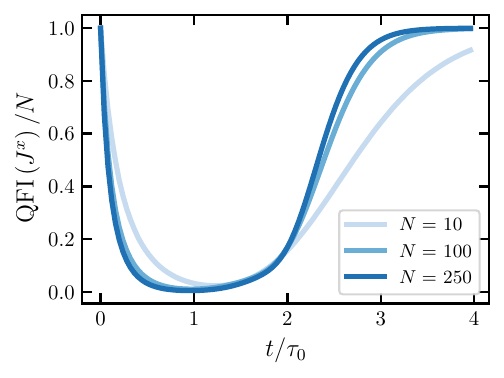}
\caption{Quantum Fisher information for the observable $J^x$ in the transient states of the super-radiant dynamics at $r_\phi = r_d =0$, and for different sizes $N$.}
\label{f.appendix_qfi}
\end{center}
\end{figure}

The quantum Fisher information (QFI) for a density matrix $\rho$ and related to an operator $\hat{O}$ is defined as \cite{pezze_quantum_2018}:
\begin{equation}
    {\rm QFI}(O) = 2 \sum_{\kappa, \kappa'} \frac{(q_{\kappa} - q_{\kappa'})^2}{q_{\kappa} + q_{\kappa'}} | \langle \kappa' | O |  \kappa \rangle |^2 
\end{equation}
where $| \kappa \rangle $ is the eigenvector of $\rho$ with eigenvalue $q_{\kappa}$. 
The maximal value of the QFI is $4 \Delta O^2$ and is reached for a pure state. 
The QFI plays a central role in quantum metrology, as it dictates the maximum sensitivity of a state to a unitary transformation generated by the operator $O$ \cite{pezze_quantum_2018}. Yet in this work we use it as an estimator of quantum coherence of the density matrix, and of the quantum contribution to the fluctuations of observables. In particular, ${\rm QFI}(O)$ measures how ``off-diagonal" the state $\rho$ is when represented in the eigenbasis of the operator $O$. 
The states $\rho$ considered in this work commute with the $J^z$ operator, hence $\rho$ is diagonal in the $J^z$ eigenbasis and ${\rm QFI}(J^z)=0$. On the other hand $[\rho, J^{x(y)}] \neq 0$, hence $\rho$ has coherences (\emph{i.e.}, off-diagonal elements) in the $J^{x(y)}$ eigenbasis. In particular, many of the $| JM\lambda\rangle$ states diagonalizing the $\rho$ states of our interest display significant quantum fluctuations of $J^{x(y)}$, particularly so when $M \approx 0$; under this condition, and if $J \sim N$,  then $\langle JM\lambda | (J^{x(y)})^2 | JM\lambda \rangle \sim O(N^2)$ (while  $\langle JM\lambda | J^{x(y)} | JM\lambda \rangle = 0$), namely the state in question has macroscopic quantum fluctuations of $J^{x(y)}$. 

Yet states $| JM\lambda\rangle$ with $M\sim 0$ are significantly populated in the steady state and in the burst dynamics when the system is in the super-radiant phase, \emph{i.e.}, when it exchanges energy very strongly with its environment and it is highly mixed. As a consequence, the high quantum coherence of each individual state of the mixture is lost in the global mixed state, and the resulting fluctuations are all of classical origin. We have already shown the absence of significant quantum fluctuations of the collective spin across the steady-state phase diagram in Sec.~\ref{s.qcorr}. 
Fig.~\ref{f.appendix_qfi} shows instead the evolution of ${\rm QFI}(J^{x(y)})$ along the super-radiant burst: we observe that the quantum coherence detected by QFI is actually minimal at the super-radiant emission peak, at which entropy reaches its maximum (see Fig.~\ref{f.dynamics_purity}). 

\end{appendix}






\bibliography{biblioSR.bib}


\end{document}